\documentclass[%
 aip,
 mph,%
 amsmath,amssymb,
 superscriptaddress,
 floatfix,
preprint,
eqsecnum,%
numerical,%
]{revtex4-1}

\usepackage{graphicx}
\DeclareGraphicsExtensions{.pdf,.png,.jpg}
\usepackage[caption=false
]{subfig}
\usepackage{epstopdf}
\usepackage{lipsum}
\usepackage{wrapfig}
\usepackage{wasysym}
\usepackage{booktabs}
\usepackage{lmodern}
\usepackage{longtable}
\usepackage{multirow}
\usepackage{amsmath}
\usepackage[OT2,T1]{fontenc}
\usepackage[bookmarks=false]{hyperref}
\usepackage{color}
\hypersetup{pdfauthor={},pdfborder=0 0 0,colorlinks=true,citecolor=blue,linkcolor=blue,urlcolor=blue,}

\DeclareSymbolFont{cyrletters}{OT2}{wncyr}{m}{n}
\DeclareMathSymbol{\comb}{\mathalpha}{cyrletters}{"58}

\newcommand{\ben}{\begin{eqnarray}\displaystyle}
\newcommand{\een}{\end{eqnarray}}

\begin{document}

\title{Deconvolution-Based Backproject-Filter (BPF) Computed Tomography Image Reconstruction Method Using Deep Learning Technique}

\author{Yongshuai Ge}
\thanks{Y. Ge and Q. Zhang are considered both as the first author.} 
\affiliation{Paul C Lauterbur Research Center for Biomedical Imaging, Research Center for Medical Artificial Intelligence, Shenzhen Institutes of Advanced Technology, Chinese Academy of Sciences, Shenzhen, Guangdong 518055, China}
\author{Qiyang Zhang}
\thanks{Y. Ge and Q. Zhang are considered both as the first author.} 
\affiliation{Paul C Lauterbur Research Center for Biomedical Imaging, Research Center for Medical Artificial Intelligence, Shenzhen Institutes of Advanced Technology, Chinese Academy of Sciences, Shenzhen, Guangdong 518055, China}
\author{Zhanli Hu}
\affiliation{Paul C Lauterbur Research Center for Biomedical Imaging, Shenzhen Institutes of Advanced Technology, Chinese Academy of Sciences, Shenzhen, Guangdong 518055, China}
\author{Jianwei Chen}
\affiliation{Department of Physics, Jinan University, Guangzhou, Guangdong 510632, China}
\author{Wei Shi}
\affiliation{Paul C Lauterbur Research Center for Biomedical Imaging, Shenzhen Institutes of Advanced Technology, Chinese Academy of Sciences, Shenzhen, Guangdong 518055, China}
\author{Hairong Zheng}
\affiliation{Paul C Lauterbur Research Center for Biomedical Imaging, Shenzhen Institutes of Advanced Technology, Chinese Academy of Sciences, Shenzhen, Guangdong 518055, China}
\author{Dong Liang}
\thanks{Scientific correspondence should be addressed to Dong Liang (dong.liang@siat.ac.cn).}
\affiliation{Paul C Lauterbur Research Center for Biomedical Imaging, Research Center for Medical Artificial Intelligence, Shenzhen Institutes of Advanced Technology, Chinese Academy of Sciences, Shenzhen, Guangdong 518055, China}

\date{\today}
 
\begin{abstract}
For conventional computed tomography (CT) image reconstruction tasks, the most popular method is the so-called filtered-back-projection (FBP) algorithm. In it, the acquired Radon projections are usually filtered first by a ramp kernel before back-projected to generate CT images. In this work, as a contrary, we realized the idea of image-domain backproject-filter (BPF) CT image reconstruction using the deep learning techniques for the first time. With a properly designed convolutional neural network (CNN), preliminary results demonstrate that it is feasible to reconstruct CT images with maintained high spatial resolution and accurate pixel values from the highly blurred back-projection image, i.e., laminogram. In addition, experimental results also show that this deconvolution-based CT image reconstruction network has the potential to reduce CT image noise (up to $20\%$), indicating that patient radiation dose may be reduced. Due to these advantages, this proposed CNN-based image-domain BPF type CT image reconstruction scheme provides promising prospects in generating high spatial resolution, low-noise CT images for future clinical applications.
\end{abstract}

\maketitle

\section{Introduction}
Until now, the FBP algorithm\cite{avinash1988principles, hsieh2009computed} and its variant the FDK cone-beam CT reconstruction algorithm\cite{feldkamp1984practical} may still be the most popular analytical CT image reconstruction method in clinical applications. In the FBP reconstruction algorithm, a 1D ramp kernel (or other type of similar filter) is usually used to filter the Radon projections, i.e.,sinogram, at first. Afterwards, the filtered sinogram is back-projected to generate the desired CT image. Since the 1D image-domain data filtration operation is simple and fast, the FBP CT image reconstruction framework has been widely used in practical implementations.

However, other analytical CT image reconstruction methods are also feasible in generating CT image in theory. For example, the Fourier slice theorem based CT image reconstruction strategy, and the backproject-filter (BPF) CT image reconstruction strategy. Herein, we mainly focus on discussing the BPF CT image reconstruction method. As the name indicated, this strategy backprojects the sinogram first to generate the laminogram\cite{smith1973image, martin1994imaging, danovich1994laminogram}. Afterwards, an image deconvolution procedure is utilized to recover the corresponding sharp CT image. To extract the true CT image, historically, one option is performing the image deconvolution in 2D Fourier domain. However, a practical difficulty of using this method roots in the fact that the ideal deconvolution kernel needs an unbounded support\cite{smith1973image}. This requires to reconstruct the laminogram onto a support considerably larger than the support of the true object, which makes the computation less efficient. Furthermore, any possible digital apodization might also degrade the spatial resolution of the final CT image. Other possible solutions could be the conventional image domain deconvolution or deblurring techniques\cite{almeida2010blind, campisi2017blind}. Despite that these approaches do not require large size support of laminogram, they may still encounter some critical challenges. For instance, the deblurring operations may be hard in recovering fine image details, or may need a prolonged image reconstruction period. Therefore, this BPF-type image reconstruction strategy was seldom implemented in history.

\begin{figure*}[!t]
\centering
\includegraphics[width=0.85\textwidth]{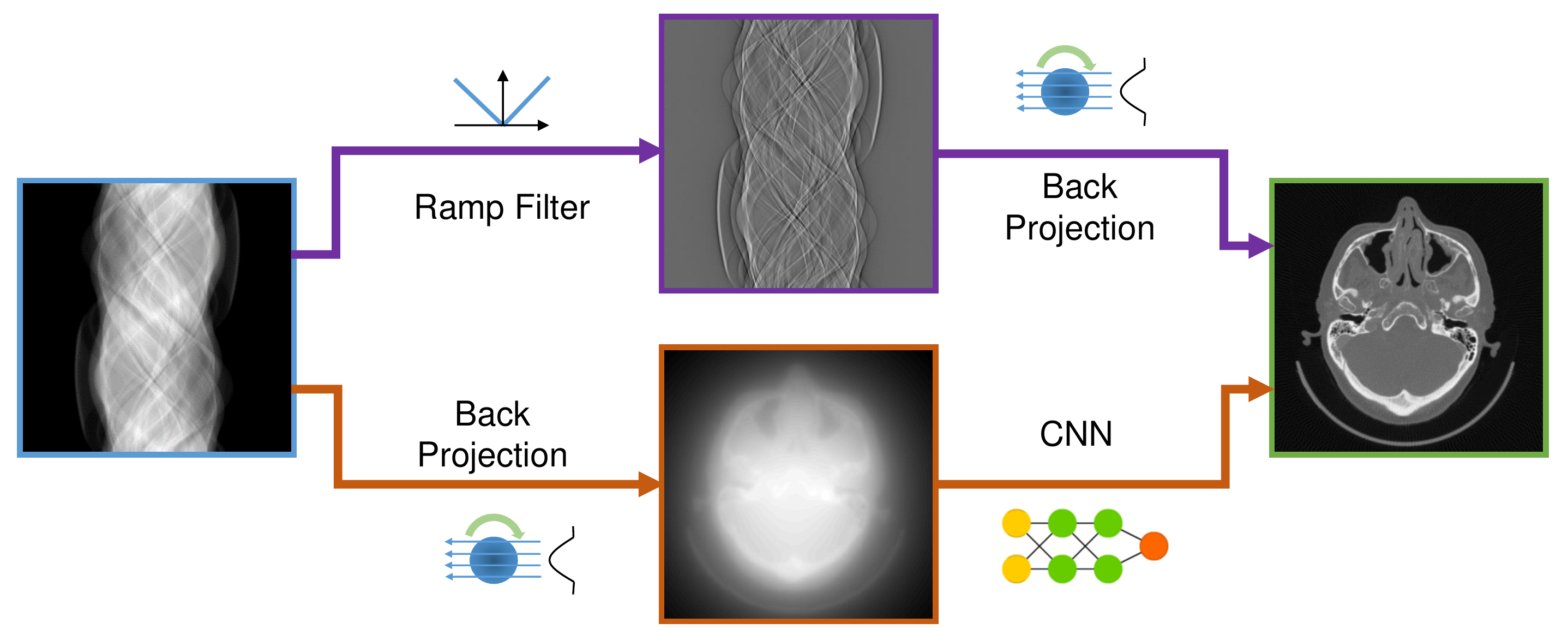}
\caption{Comparison of two CT image reconstruction schemes: the top one represents the conventional FBP reconstruction strategy; the bottom one represents the proposed CNN-based BPF reconstruction strategy.}
\label{fig:sino_ed} 
\end{figure*}

Nowadays, the fast developing deep learning technique has shown great success in image processing fields, including classification, identification, and segmentation\cite{lecun2015deep}. Definitely, this new emerging technique will provide many opportunities for medical CT imaging applications. Some of the latest CT image processing studies\cite{jin2017deep, chen2017low,wang2018image, han2018framing,kang2018deep,yang2018low,zhang2018sparse} have demonstrated the advancements of the deep learning technique, especially the deep convolutional neural network (CNN), in reducing image noise and removing image artifacts. In one of the latest studies, it has also demonstrated that CNN can learn the CT image reconstruction procedures via domain transform manifold learning\cite{zhu2018image}. These exciting and pioneering studies strongly indicate that the deep learning technique can help solving problems hard to be solved in history, and providing more efficient and better solutions than conventionally used methods. Under this condition, in this paper, we proposed to reconstruct CT image by deconvolving the highly blurred laminogram using the latest CNN technique. In particular, we would like to investigate the feasibility of reconstructing CT image from the laminogram via image-domain deconvolution using the CNN technique. The workflow of the conventional FBP reconstruction method and the new CNN-based BPF reconstruction method are compared in Fig.~\ref{fig:sino_ed}.

To our best knowledge, it is the first time to revisit the BPF-type CT image reconstruction idea using deep learning technique. This preliminary work mainly focuses on demonstrating the feasibility of this newly suggested strategy. Other advancements of this CNN-based new CT image reconstruction framework, for example, the CT image denoising\cite{chen2017low,kang2018deep,yang2018low} capability and the CT image artifact removal\cite{ han2018framing,zhang2018sparse} capability, will be studied and discussed in future. The main contributions of this work are as follows: First, we demonstrated that a properly designed CNN architecture is able to learn a close approximation of the theoretical deconvolution kernel, and therefore can reconstruct the object from laminogram with high accuracy; Second, the reconstructed CT image has similar spatial resolution compared with the FBP reconstructed CT image; Third, CT images generated from this CNN-based BPF reconstruction method also maintains similar noise texture as in the FBP reconstruction method. Fourth, the CNN-based BPF reconstruction method has the potential to reduce CT image noise (up to $20\%$) compared with the FBP results. To prove these observations, both numerical and experimental studies are conducted, results and quantitative measurements are presented in the following sections as well.

\section{Methods}
\subsection{Theoretical foundation}
Without loss of generality, in this theoretical discussion section, we assume that the CT system has a parallel-beam imaging geometry. Similar conclusions may also be obtained for the fan-beam geometry\cite{gullberg1995backprojection}. For parallel-beam CT imaging geometry, the projection of an arbitrary 2D object $f(x,y)$ on the detector plane at a certain view-angle $\theta$ can be written as following:
\begin{equation}
p(r,\theta) = \int^{+\infty}_{-\infty}\int^{+\infty}_{-\infty} f(x,y) \delta (x\cos \theta+y\sin \theta-r) \mathrm{d} x \mathrm{d} y.
\label{eq:parallel_radon}
\end{equation}
In Eq.~(\ref{eq:parallel_radon}), $p(r,\theta)$ denotes the measured projection, $\delta(x\cos \theta+y\sin \theta-r)$ specifies an x-ray beam passing through the object, $r$ is the distance from the origin $(x=0,y=0)$ to the beam, $\theta$ is the angle between the $+y$-axis to the ray penetrating the object. For the parallel-beam imaging geometry, the projections are usually acquired with 180 degree tube-detector rotation interval, namely, $0\leq\theta\leq\pi$. Usually, Eq.~(\ref{eq:parallel_radon}) is also known as the Radon transform\cite{avinash1988principles, hsieh2009computed}.

To continue the following discussions, now we simply take each projection value and put it back into the object space along the Radon integration direction one view after another. Thus, the so-called laminogram, denoted as $f_{b}(x,y)$, is generated. Explicitly, this angularly-weighted back-projection procedure can be expressed as:
\begin{equation}
f_{b}(x,y) = \int^{\pi}_{0}p(r,\theta) \delta (r-x\cos \theta-y\sin \theta) \mathrm{d} \theta.
\label{eq:bp}
\end{equation}
By substituting Eq.~(\ref{eq:parallel_radon}) into Eq.~(\ref{eq:bp}), one can get the following relationship between the laminogram image $f_{b}(x,y)$ and the reference object $f(x,y)$,
\begin{equation}
f_{b}(x,y) = f(x,y)\otimes\frac{1}{\sqrt{(x^2+y^2)}}.
\label{eq:bp_1}
\end{equation}
Evidently, the direct back-projection of the Radon projections generates a blurred version of the reference object $f(x,y)$. The blurry kernel is $1/\sqrt{(x^2+y^2)}$. The 2D Fourier transformation of Eq.~(\ref{eq:bp_1}) yields,
\begin{equation}
\mathcal{F}_{b}(u,v) = \mathcal{F}(u,v)\times\frac{1}{\sqrt{(u^2+v^2)}},
\label{eq:bp_2}
\end{equation}
where variables $(u,v)$ are the Fourier domain counterparts to the image domain spatial variables $(x,y)$. Note that the convolution operation in image-domain as shown in Eq.~(\ref{eq:bp_1}) has replaced by the multiplication operation in frequency-domain, as shown in Eq.~(\ref{eq:bp_2}). Immediately, one can get,
\begin{equation}
\mathcal{F}(u,v) = \mathcal{F}_{b}(u,v)\times\sqrt{(u^2+v^2)}.
\label{eq:bp_3}
\end{equation}
Upon Eq.~(\ref{eq:bp_3}), now we can recover the reference object $f(x,y)$ by taking an inverse Fourier transform on both sides,
\begin{align}
f(x,y)&=\mathrm{FT}^{-1}\{\mathcal{F}(u,v)\},\\
& = \mathrm{FT}^{-1}\{\mathcal{F}_{b}(u,v)\times\sqrt{(u^2+v^2)}\}.
\label{eq:bp_4}
\end{align}
This is one possible solution, and potential difficulties of solving it has been briefly discussed before. Actually, according to the convolution theorem, the product of the Fourier transforms of two functions in frequency space equals to the convolution of two functions in spatial space. Therefore, Eq.~(\ref{eq:bp_4}) can be rewritten as below,
\begin{equation}
f(x,y) = f_{b}(u,v)\otimes\mathrm{FT}^{-1}\{\sqrt{(u^2+v^2)}\}.
\label{eq:bp_5}
\end{equation}
The Eq.~(\ref{eq:bp_5}) provides another option in reconstructing the reference object $f(x,y)$. In this approach, deconvolution is performed in image space directly. As mentioned above, conventional image domain deconvolution algorithms may encounter some difficulties. Therefore, the deep learning technique will be introduced and investigated to learn an approximation of the ideal deconvolution kernel $\mathrm{FT}^{-1}\{\sqrt{(u^2+v^2)}\}$ in Eq.~(\ref{eq:bp_5}).

\begin{figure}[h]
\centering
\includegraphics[width=0.35\textwidth]{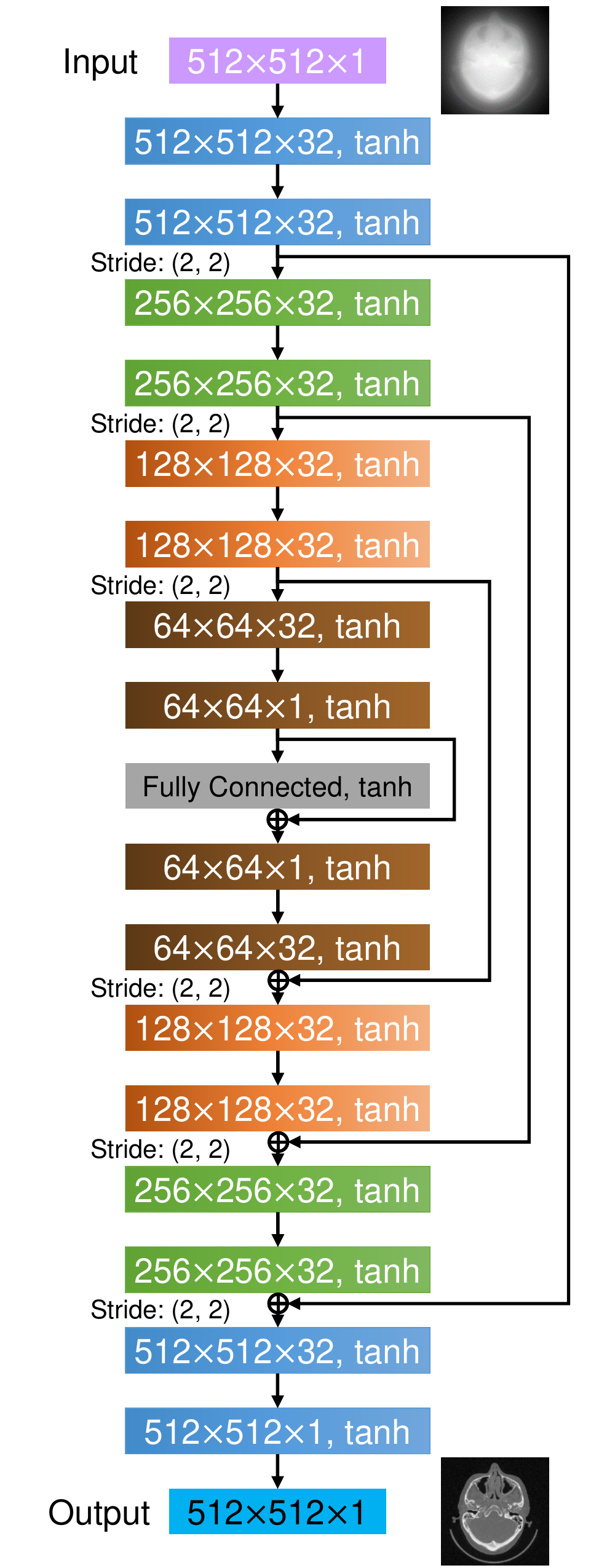}
\caption{The architecture of the CNN network. In total there are 17 individual layers. The weight kernels for the convolutional layers are $3\times3$ except $1\times1$ for the last convolutional layer. The middle of this CNN is a fully-connected layer. The three numbers in each box denotes the image column number, image row number, and the channel number, respectively.  All activation functions used are tanh. Four shortcuts are added symmetrically with respect to the fully-connected layer.}
\label{fig:network_ed} 
\end{figure}

\subsection{Convolutional neural network architecture}
In this study, the convolutional neural network feed-forward architecture was used. The CNN extracts high-level features from the blurred laminogram, and forces the entire network to approximate the theoretical deconvolution kernel. The input of the network is laminogram, the output is the deconvolved CT image, and between the input and the output is the network, as shown in Fig.~\ref{fig:network_ed}. The designed architecture contains 17 layers in total: the first eight convolutional layers, the fully-connected layer in middle, and the last eight convolutional layers. The first eight convolutional layers help to encode useful image information from the laminogram. During this procedure, three stride=(2, 2) operations are used to reduce the image size from $512\times512$ down to $64\times64$. The last eight convolutional layers after the fully-connected layer form the information decoder and generate the expected sharp CT images. As a contrary to the encoder, the image is upsampled step by step from $64\times64$ back to the original $512\times512$ image size. The weight kernels for the convolutional layers are $3\times3$ except $1\times1$ for the last convolutional layer. By design, four shortcuts are added accordingly to minimize the vanishing gradient phenomenon during back-propagation\cite{he2016deep}.

The network objective loss function is defined as:
\begin{equation}
\mathcal{L} = \frac{1}{N_x}\frac{1}{N_y}\sum^{N_x}_{i=1}\sum^{N_y}_{j=1}|I(i,j)-I_{cnn}(i,j)|^2,
\label{eq:loss}
\end{equation}
where $N_x$ and $N_y$ represent the total pixels of CT image along the horizontal and vertical directions, correspondingly. In addition, the reference CT image is denoted as $I$, the CNN-learned CT image is denoted as $I_{cnn}$, and a pair of $(i, j)$ represents a certain pixel on image. During training procedures, the network automatically updates the weights and biases for all channels, and gradually learns the optimal parameters to minimize the loss function $\mathcal{L}$. As the training continues, the learned CT image $I_{cnn}$ converges to the reference image $I$. Specifically, the Adam algorithm\cite{kingma2014adam} was used with starting learning rate of 0.0003. The learning rate was exponentially decayed by a factor of 0.96 after every 1000 steps. The mini-batch had a size of 16, and batch-shuffling was turned on to increase the randomness of the training data. The network was trained for 100 epochs on the Tensorflow deep learning framework using a single graphics processing unit (GPU, NVIDIA GeForce GTX 1080Ti) with 11 GB memory capacity.


\subsection{Numerical study: Materials and methods}
To verify the feasibility of this novel CNN based CT image reconstruction algorithm, numerical studies were performed with 50,000 diagnostic CT images downloaded from The Cancer Imaging Archive (TCIA). Anatomically, these diagnostic CT images contain typical human body structures: head and neck, chest, abdomen, and pelvis. All CT images were reconstructed into a $512\times512$ image matrix by vendors. 

To build the training dataset, a linear transformation was applied to convert the original CT image value of Hounsfield Unit (HU) to the linear attenuation coefficient $\mu$ as following:
\begin{equation}
\mu = 0.02\times \left \{ \frac{\mathrm{HU}}{1000}+1\right\},
\label{eq:hu_mu}
\end{equation}
where the factor 0.02 (with unit of 1/mm) corresponds to the reference X-ray linear attenuation coefficient of water $\mu^{ref}_{water}$ around 60 keV. This is a close approximation to most of the mean X-ray tube energy in routine clinical CT scans. Essentially, this conversion also helps to normalize the TICA CT images. To augment the dataset, CT images were rotated by 90 degrees randomly. In addition, a uniform random number distributed between 0.50 to 2.00 was further multiplied to the CT images to broaden the data range. 

A fan-beam CT imaging geometry was simulated. The source to detector distance was 1500.00 mm, and the source to rotation center was 1000.00 mm. There were 1400 detector elements, each had a dimension of 0.30 mm. To make the clinical CT images fit in this simulated imaging geometry, we further assumed that all CT images have the same pixel dimension of $0.35 mm \times 0.35 mm$. Forward projections, i.e., Radon projections, were collected from 360 views with 1.00 degree angular interval. Notice that this simulated fan-beam CT imaging geometry is different from most of the real multi-slice diagnostic CT (MDCT) scanner imaging geometry. Also be aware that no noise is added in these simulations.

\begin{figure}[h]
\centering
\includegraphics[width=0.35\textwidth]{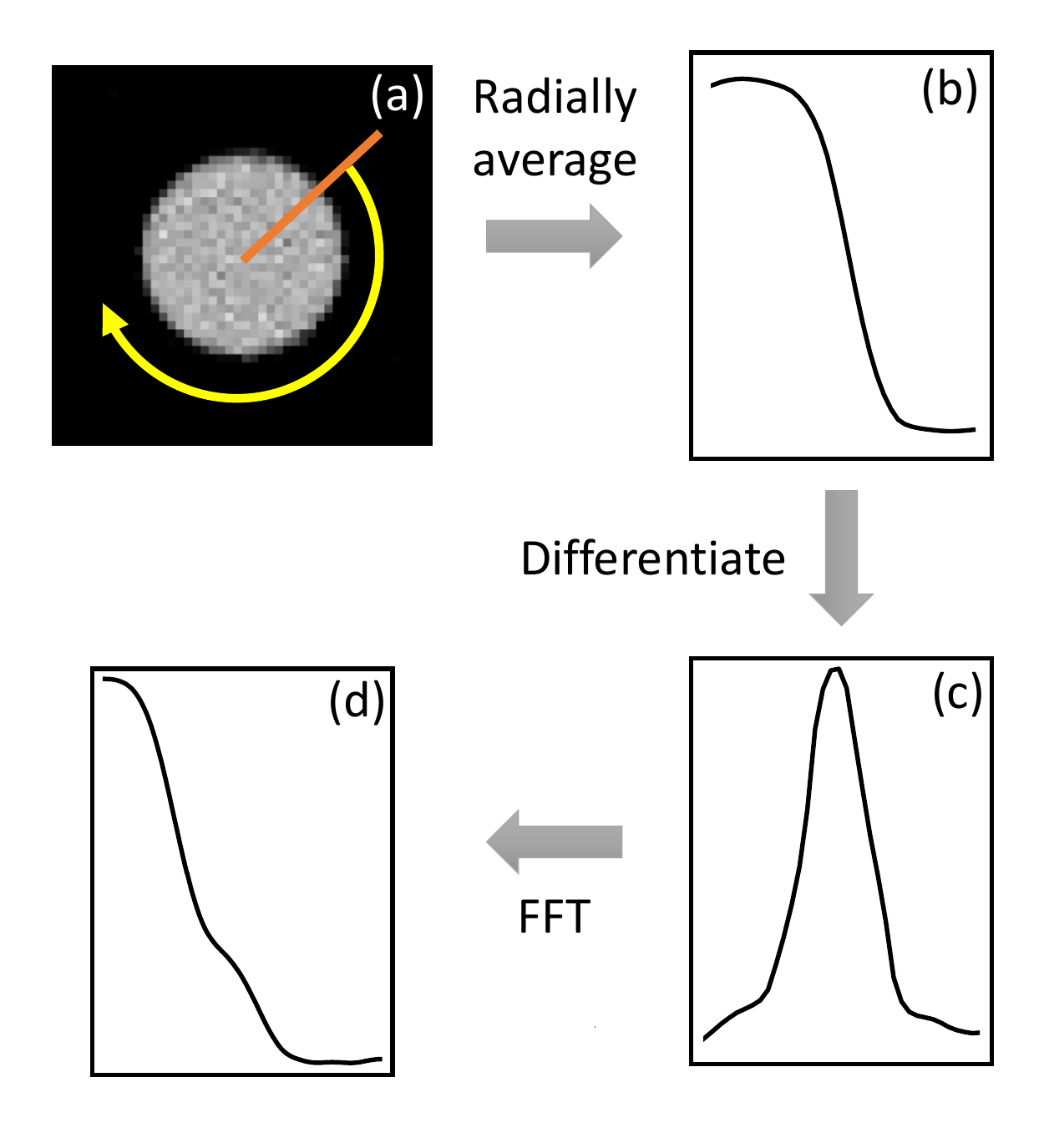}
\caption{Flowchart to calculate MTF curves: (a) the cropped image with the Teflon rod insert positioned in center, (b) the radially averaged edge spread function, (c) the corresponding line spread function, (d) the estimaged MTF curve.}
\label{fig:mtf_ed} 
\end{figure}

\begin{figure*}[h]
\centering
\includegraphics[width=0.95\textwidth]{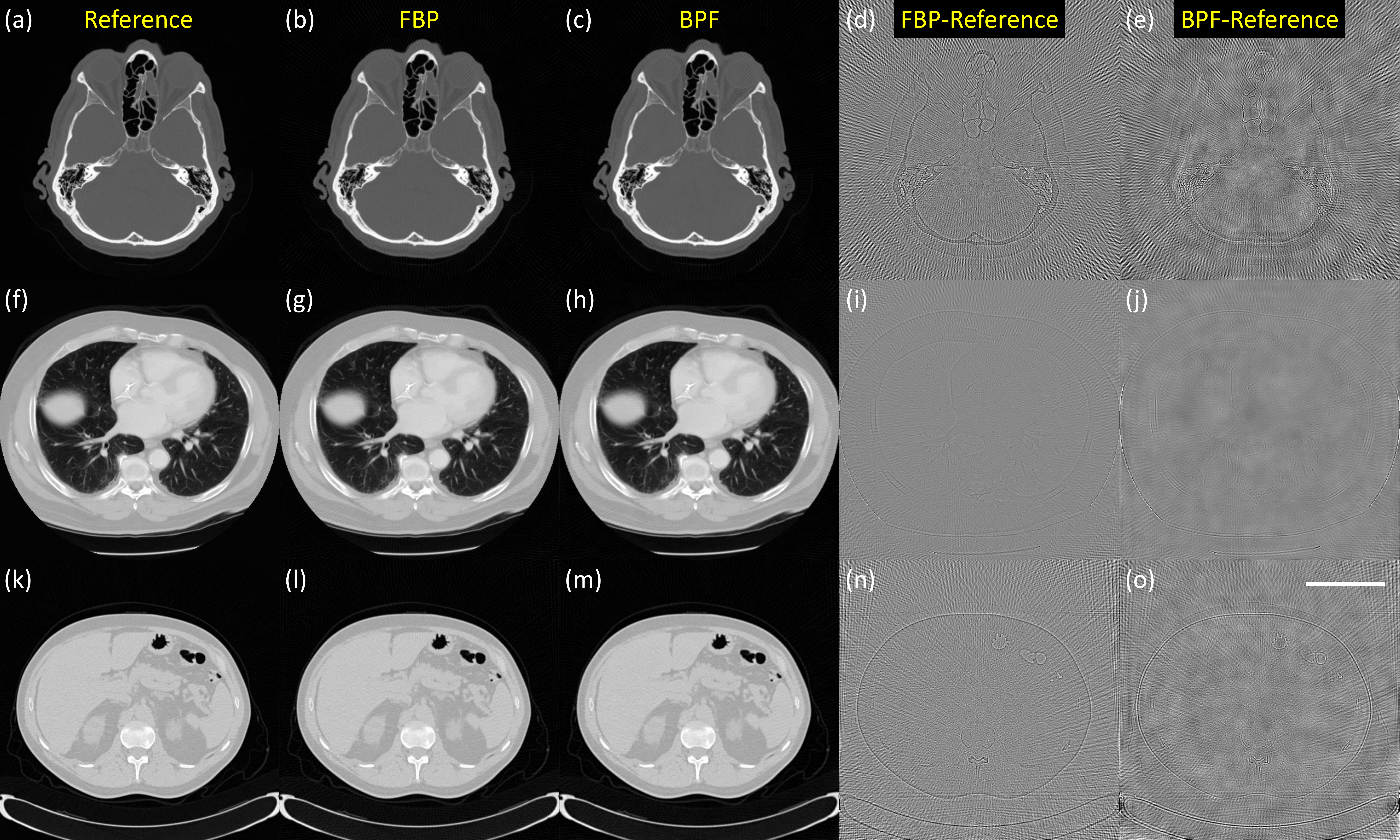}
\caption{Validation results using clinical CT images obtained from TCIA. The display window is [0.00, 0.04] for the CT images, and is [-0.003, 0.003] for the difference images. The scale bar denotes 5.00 cm. }
\label{fig:Num_results} 
\end{figure*}

Afterwards, laminograms with $512\times512$ image matrix size were reconstructed by directly back-projects the sinograms without adding any filtration. Thus, one laminogram (as the network input) and one reference CT image (as the network label) were prepared and paired to form both the training and testing datasets. Overall, 50,000 such data pairs were collected: 49,700 pairs were generated to form the training dataset, and the rest 300 pairs were prepared as the testing dataset. All these above numerical operations were performed in Matlab (The MathWorks Inc., Natick, MA, USA), as well as Matlab-executable CUDA (Compute Unified Device Architecture) algorithms, which are responsible for the forward Radon projections and the back-projections.

\subsection{Experimental materials and methods}
Experimental validations were further conducted to demonstrate the performance and robustness of the proposed BPF CT image reconstruction framework. All experiments were performed on an in-house CT imaging bench in our lab. The system had a rotating-anode Tungsten target diagnostic level X-ray tube (Varex G-242, Varex Imaging Corporation, UT, USA). It was operated at 120.00 kV continuous fluoroscopy mode with 0.40 mm nominal focal spot. The X-ray tube current was set at 11.00 mA. The collimated beam had a vertical height of 25.00 mm at the rotation-center. A bowtie filter designed for PMMA phantom with 320.00 mm diameter was used, together with additional 0.50 mm Copper filtration. The X-ray detector was an energy-resolving photon-counting detector (XC-Hydra FX50, XCounter AB, Sweden) made from CdTe (Cadmium Telluride) material. It had an imaging area of 512.00 mm $\times$ 6.00 mm with a native element dimension of 0.10 mm $\times$ 0.10 mm. The detector signal accumulation period was 0.10 second. Projection datasets were acquired with $1\times1$ detector binning mode, and were $6\times6$ rebinned in the post-processing procedures. Since there was no need to discriminate photon energies, the detector was operated and calibrated\cite{ge2017k} with only one single energy threshold (10.00 keV). To enable more accurate photon number detection, its anti-charge-sharing function was turned on. The source to detector distance was equal to 1500.00 mm, and the source to rotation-center was 1000.00 mm. A CatPhan-700 (The Phantom Laboratory Inc., Salem, NY, USA), and a dental and diagnostic head phantom (Atom Max 711-HN, CIRS Inc., VA, USA) were imaged in this work. Phantoms were rotated by 360.00 degrees with 1.00 degree angular interval. Under this particular system settings, the measured $\mathrm{CTDI_{vol}}$ from a PMMA phantom of 160.00 mm diameter was 2.00 mGy. To investigate the radiation dose dependency of the new BPF-type CT image reconstruction algorithm, phantoms were scanned repeated by 10 times. Thus, the CT images generated with minimum noise correspond to a $\mathrm{CTDI_{vol}}$ of 20.00 mGy. No beam-hardening corrections and scatter corrections were performed during these experiments.

\begin{figure*}[h]
\centering
\includegraphics[width=0.90\textwidth]{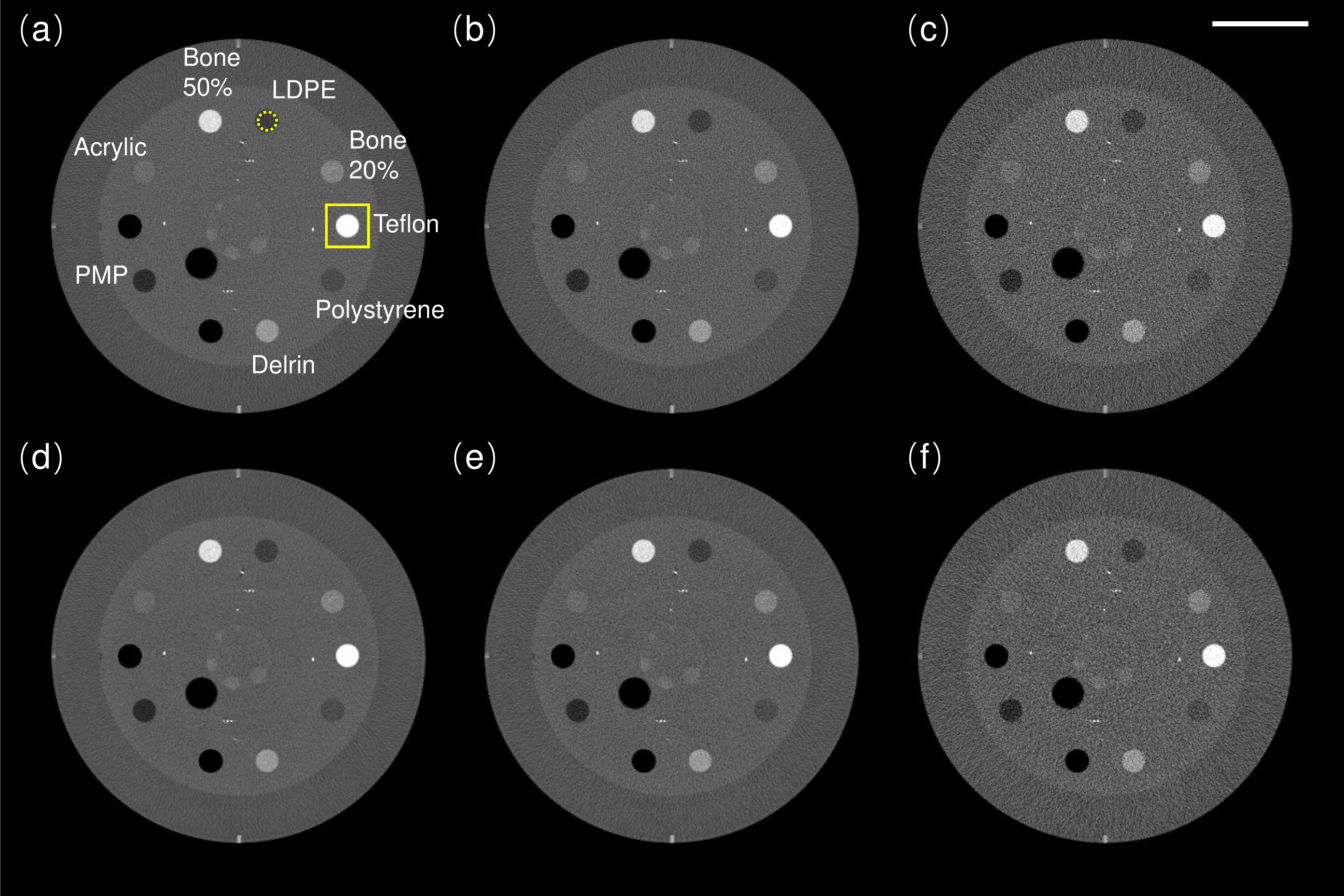}
\caption{Experimental results of the CTP682 module in CatPhan-700. Images in (a)-(c) are reconstructed from the FBP algorithm, and images in (d)-(f) are obtained from the BPF algorithm. Images in (a) and (d) are acquired with $100\%$ radiation dose; Images in (b) and (e) are acquired with $50\%$ radiation dose; Images in (c) and (f) are acquired with $10\%$ radiation dose. The scale bar denotes 5.00 cm. The display window is [0.01, 0.03].}
\label{fig:Exp_results_1_01} 
\end{figure*}

Experimental CT images were reconstructed readily via two different approaches: one is the conventional FBP method; the other is the newly developed CNN deconvolution-based method in this paper. For the first FBP approach, the filtration kernel was a standard ramp kernel. For the second BPF approach, sinogram was first backprojected to get the laminogram, and then immediately put into the CNN network to reconstruct the desired CT image. The used CNN parameters were the same as used for the numerical validations in previous section. Both the laminogram and the reconstructed CT image have size of $512\times512$. The voxel dimension of CT images for visualization is $0.50 mm\times0.50 mm\times0.50 mm$. The voxel dimension was decreased to $0.35 mm\times0.35 mm\times0.35 mm$ for quantitative image evaluations. Finally, the signal accuracy, image spatial resolution, i.e., the modulation transfer function (MTF), and the noise power spectrum (NPS) were studied side by side for the two methods. The workflow for MTF calculation is illustrated in Fig.~\ref{fig:mtf_ed}.

\section{Results}
\subsection{Results of numerical simulations}
In this section, the numerical validations are presented. Results are illustrated in Fig.~\ref{fig:Num_results}. In it, the first column represents the reference CT images, the second column represents the standard FBP-type CT images reconstructed from the acquired 360 views of projections, the third column represents the CNN learned BPF-type CT images, the fourth and fifth columns represent the difference between the reference and the reconstructed images correspondingly. Clearly, the proposed CNN network is able to reconstruct sharp CT images from the highly blurred laminograms with very high accuracy. 

\subsection{Results of experimental studies}
\begin{figure*}[t]
\centering
\includegraphics[width=0.90\textwidth]{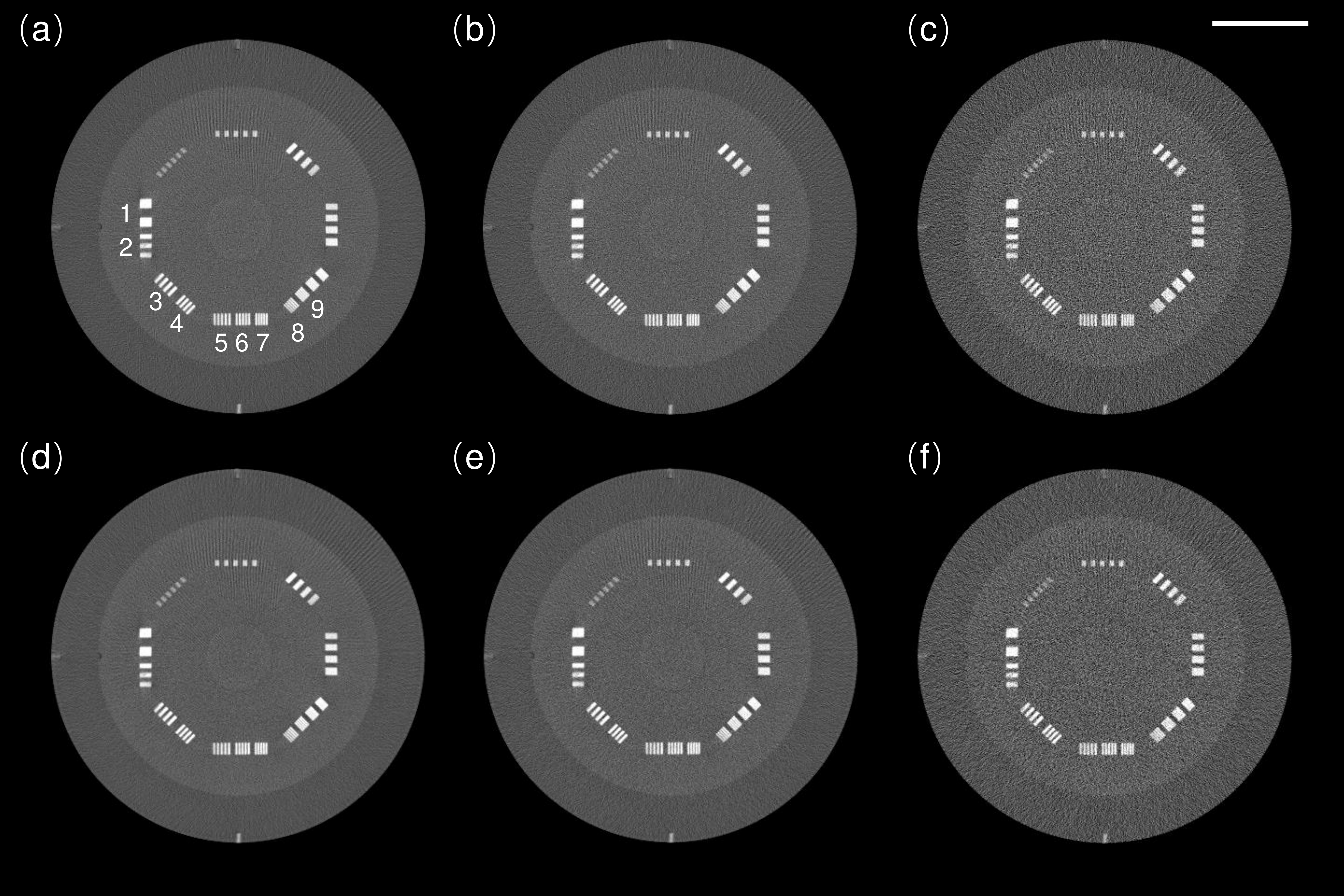}
\caption{Experimental results of the CTP714 module in CatPhan-700. Images in (a)-(c) are reconstructed from the FBP algorithm, and images in (d)-(f)  are obtained from the BPF algorithm. Images in (a) and (d) are acquired with $100\%$ radiation dose; Images in (b) and (e) are acquired with $50\%$ radiation dose; Images in (c) and (f) are acquired with $10\%$ radiation dose. The scale bar denotes 5.00 cm. The display window is [0.01, 0.03].}
\label{fig:Exp_results_1_02} 
\end{figure*}

\begin{figure*}[h]
\centering
\includegraphics[width=0.90\textwidth]{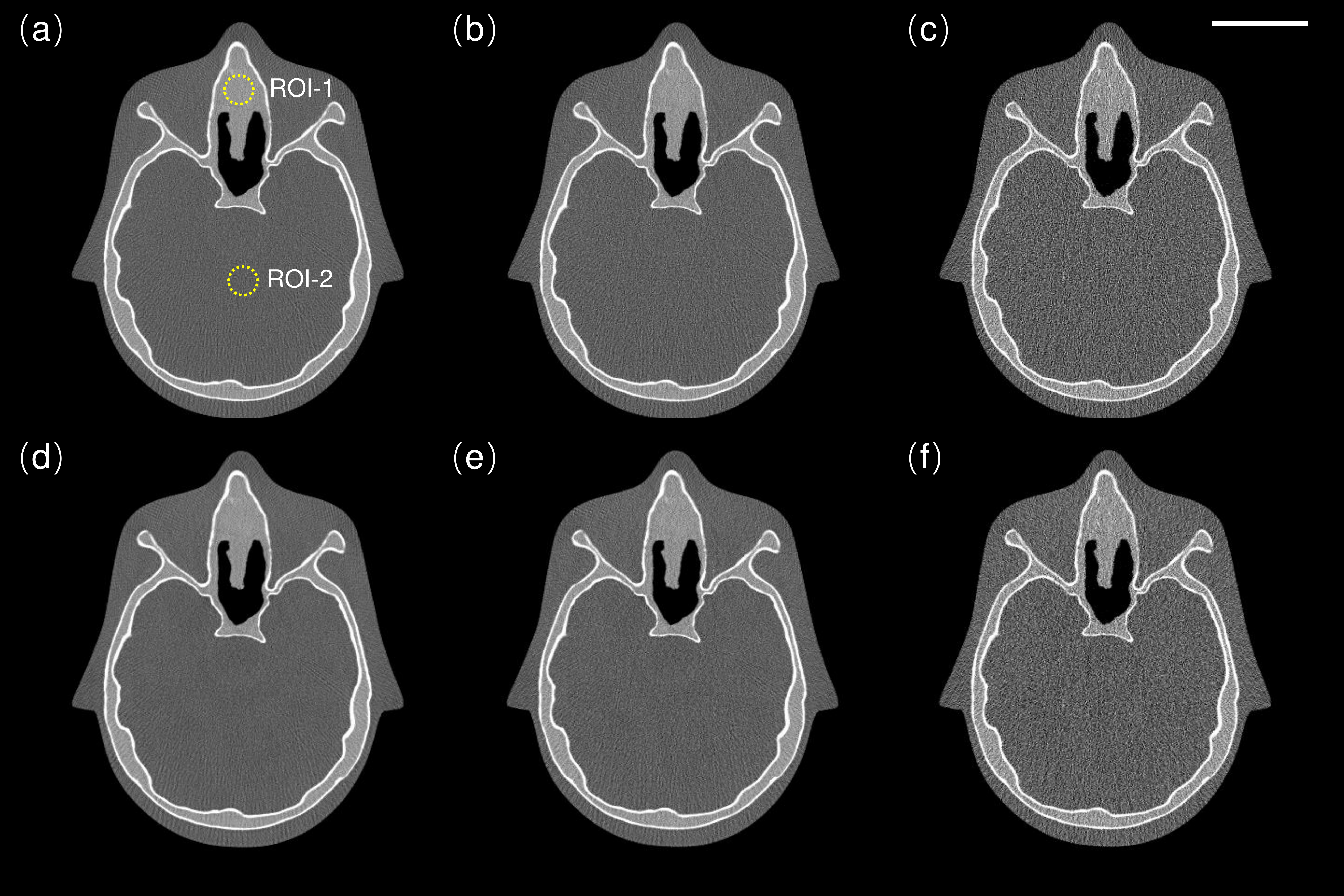}
\caption{Experimental results of the head and neck phantom. Images in (a)-(c) are reconstructed from the FBP algorithm, and images in (d)-(f) are obtained from the BPF algorithm. Images in (a) and (d) are acquired with $100\%$ radiation dose; Images in (b) and (e) are acquired with $50\%$ radiation dose; Images in (c) and (f) are acquired with $10\%$ radiation dose. The scale bar denotes 5.00 cm. The display window is [0.01, 0.03].}
\label{fig:Exp_results_1_03} 
\end{figure*}

Experimental results are presented in this section. Figs.~\ref{fig:Exp_results_1_01}-\ref{fig:Exp_results_1_03} compare the reconstructed CT images from both the FBP method and the CNN-based BPF method for three different radiation dose levels: $100\%$ radiation dose; $50\%$ radiation dose; and $10\%$ radiation dose. Since there are no reference images available, the standard FBP reconstructed CT images are considered as the ground truth, and the CT images reconstructed from the CNN network are compared with them. 

The imaging results include two different internal modules inside the CatPhan-700 phantom, and one central slice of the head and neck phantom. To enable quantitative comparisons of the signal values, ten ROIs (region-of-interests) are selected to measure their mean signal values and the standard deviations. Results are listed in Table~\ref{table:1}. Herein, the first eight ROIs are obtained from the rod inserts inside CatPhan-700 phantom, as marked on Fig.~\ref{fig:Exp_results_1_01}(a). The last two ROIs are drawn on the head and neck phantom, as highlighted on Fig.~\ref{fig:Exp_results_1_03}(a). These measurements demonstrate that the CNN-based BPF-type CT image reconstruction method is able to generate CT images with accurate signal values, and has the potential to reduce image noise as well. Take the measured results of LDPE rod for example, the relative difference of the mean signal value obtained from the BPF reconstruction method ($1.72\times10^{-2}$) to the FBP method ($1.73\times10^{-2}$) is within $\pm1\%$. Meanwhile, the relative difference of the signal standard deviation obtained from the BPF reconstruction method ($1.02\times10^{-3}$) to the FBP method ($1.22\times10^{-3}$) is up to about $20\%$.

\begin{table*}[t]
\centering
\caption{Measured mean values and standard deviations from ten different ROIs (Unit: $1/mm$).}
\label{table:1}
  \begin{tabular}{c|cc|cc|cc}
    \hline
    \hline
    {Dose} &
  \multicolumn{2}{c|}{100\%} &
  \multicolumn{2}{c|}{50\%} &
  \multicolumn{2}{c}{10\%} \\
   \hline
    {Algorithm} & {\quad\quad FBP  \quad\quad} & {\quad\quad BPF \quad} & {\quad\quad FBP  \quad} & {\quad\quad BPF \quad\quad} & {\quad\quad FBP  \quad} & {\quad\quad BPF \quad\quad} \\
      \hline
    Bone 50\% & 3.14$\times10^{-2}\pm$ & 3.13$\times10^{-2}\pm$ & 3.14$\times10^{-2}\pm$ & 3.14$\times10^{-2}\pm$ & 3.14$\times10^{-2}\pm$ & 3.13$\times10^{-2}\pm$ \\[-0pt]
    & 9.40$\times10^{-4}$ & 8.30$\times10^{-4}$ & 1.23$\times10^{-3}$ & 1.06$\times10^{-3}$ & 2.56$\times10^{-3}$ & 2.10$\times10^{-3}$  \\[3pt]

    LDPE & 1.73$\times10^{-2}\pm$ & 1.73$\times10^{-2}\pm$ & 1.73$\times10^{-2}\pm$ & 1.72$\times10^{-2}\pm$ & 1.73$\times10^{-2}\pm$ & 1.73$\times10^{-2}\pm$ \\[-0pt]
    & 9.20$\times10^{-4}$ & 7.70$\times10^{-4}$ & 1.22$\times10^{-3}$ & 1.02$\times10^{-3}$ & 2.52$\times10^{-3}$ & 2.10$\times10^{-3}$  \\[3pt]

    Bone 20\% & 2.33$\times10^{-2}\pm$ & 2.32$\times10^{-2}\pm$ & 2.32$\times10^{-2}\pm$ & 2.32$\times10^{-2}\pm$ & 2.33$\times10^{-2}\pm$ & 2.31$\times10^{-2}\pm$ \\[-0pt]
    & 1.14$\times10^{-3}$ & 9.80$\times10^{-4}$ & 1.43$\times10^{-3}$ & 1.21$\times10^{-3}$ & 2.75$\times10^{-3}$ & 2.31$\times10^{-3}$  \\[3pt]

    Teflon & 3.58$\times10^{-2}\pm$ & 3.57$\times10^{-2}\pm$ & 3.58$\times10^{-2}\pm$ & 3.57$\times10^{-2}\pm$ & 3.58$\times10^{-2}\pm$ & 3.58$\times10^{-2}\pm$ \\[-0pt]
    & 1.01$\times10^{-3}$ & 9.10$\times10^{-4}$ & 1.36$\times10^{-3}$ & 1.17$\times10^{-3}$ & 2.89$\times10^{-3}$ & 2.37$\times10^{-3}$  \\[3pt]

    PS & 1.84$\times10^{-2}\pm$ & 1.85$\times10^{-2}\pm$ & 1.84$\times10^{-2}\pm$ & 1.85$\times10^{-2}\pm$ & 1.84$\times10^{-2}\pm$ & 1.85$\times10^{-2}\pm$ \\[-0pt]
    & 9.20$\times10^{-4}$ & 7.60$\times10^{-4}$ & 1.29$\times10^{-3}$ & 1.09$\times10^{-3}$ & 2.57$\times10^{-3}$ & 2.21$\times10^{-3}$  \\[3pt]

    Delrin & 2.53$\times10^{-2}\pm$ & 2.51$\times10^{-2}\pm$ & 2.53$\times10^{-2}\pm$ & 2.51$\times10^{-2}\pm$ & 2.53$\times10^{-2}\pm$ & 2.51$\times10^{-2}\pm$ \\[-0pt]
    & 8.50$\times10^{-4}$ & 7.10$\times10^{-4}$ & 1.24$\times10^{-3}$ & 1.04$\times10^{-3}$ & 2.72$\times10^{-3}$ & 2.26$\times10^{-3}$  \\[3pt]

    PMP & 1.58$\times10^{-2}\pm$ & 1.55$\times10^{-2}\pm$ & 1.57$\times10^{-2}\pm$ & 1.55$\times10^{-2}\pm$ & 1.58$\times10^{-2}\pm$ & 1.56$\times10^{-2}\pm$ \\[-0pt]
    & 7.70$\times10^{-4}$ & 6.40$\times10^{-4}$ & 1.06$\times10^{-3}$ & 9.00$\times10^{-4}$ & 2.38$\times10^{-3}$ & 2.07$\times10^{-3}$  \\[3pt]

    Acrylic & 2.12$\times10^{-2}\pm$ & 2.10$\times10^{-2}\pm$ & 2.11$\times10^{-2}\pm$ & 2.09$\times10^{-2}\pm$ & 2.11$\times10^{-2}\pm$ & 2.09$\times10^{-2}\pm$ \\[-0pt]
    & 9.90$\times10^{-4}$ & 8.20$\times10^{-4}$ & 1.32$\times10^{-3}$ & 1.11$\times10^{-3}$ & 2.45$\times10^{-3}$ & 2.08$\times10^{-3}$  \\[3pt]
    
    ROI-1 & 2.50$\times10^{-2}\pm$ & 2.48$\times10^{-2}\pm$ & 2.50$\times10^{-2}\pm$ & 2.48$\times10^{-2}\pm$ & 2.51$\times10^{-2}\pm$ & 2.50$\times10^{-2}\pm$ \\[-0pt]
    & 1.22$\times10^{-3}$ & 1.03$\times10^{-3}$ & 1.59$\times10^{-3}$ & 1.33$\times10^{-3}$ & 3.29$\times10^{-3}$ & 2.64$\times10^{-3}$  \\[3pt]

    ROI-2 & 2.00$\times10^{-2}\pm$ & 1.98$\times10^{-2}\pm$ & 2.00$\times10^{-2}\pm$ & 1.98$\times10^{-2}\pm$ & 2.00$\times10^{-2}\pm$ & 1.97$\times10^{-2}\pm$ \\[-0pt]
    & 1.20$\times10^{-3}$ & 1.00$\times10^{-3}$ & 1.64$\times10^{-3}$ & 1.34$\times10^{-3}$ & 3.64$\times10^{-3}$ & 2.92$\times10^{-3}$  \\
    \hline
    \hline
  \end{tabular}
\end{table*}

\begin{figure}[t]
\centering
\includegraphics[width=0.90\textwidth]{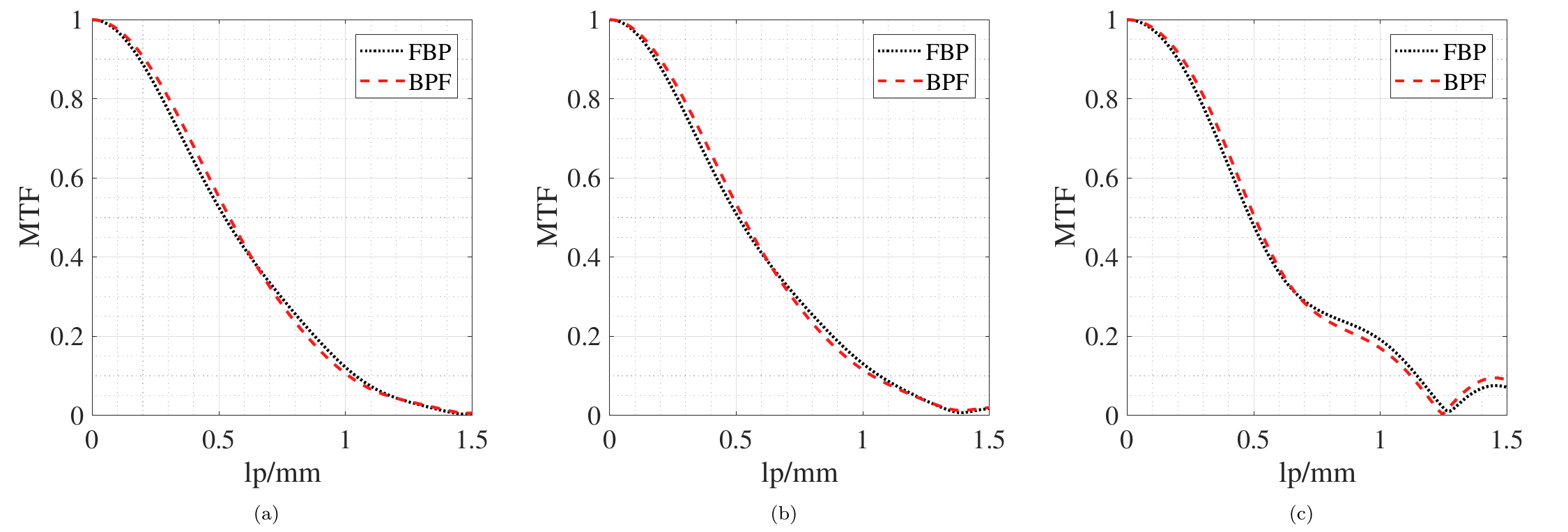}
\vspace{-0.1in}
\caption{The measured MTF curves from the Teflon rod at three different dose levels: (a) $100\%$ radiation dose, (b) $50\%$ radiation dose, (c) $10\%$ radiation dose.}
\label{fig:mtf_01} 
\end{figure}

Moreover, the MTFs are further measured with the Teflon rod (see the squared region on Fig.~\ref{fig:Exp_results_1_01}(a)) in the CatPhan-700 phantom at three different radiation dose levels for both the FBP method and the BPF method. Detailed signal processing procedures can be found in Fig.~\ref{fig:mtf_ed}. The MTF results are shown in Fig.~\ref{fig:mtf_01}. Overall, the CNN-based BPF method generates similar MTFs as the conventional FBP method. In particular, the BPF reconstruction method generates slightly better MTF response at low spatial frequency range than the FBP reconstruction method. However, the FBP reconstruction method slightly outperforms the BPF reconstruction method at high frequency range. Additionally, the imaging results of resolution bar structures in Fig.~\ref{fig:Exp_results_1_02} enable to visually appreciate the similar spatial resolution performance of the FBP and BPF reconstruction algorithms. For all radiation dose levels, the bar structures of the seventh group can be clearly distinguished for both reconstruction methods, which agree well with the previous MTF quantification results.

\begin{figure*}[h]
\centering
\includegraphics[width=0.90\textwidth]{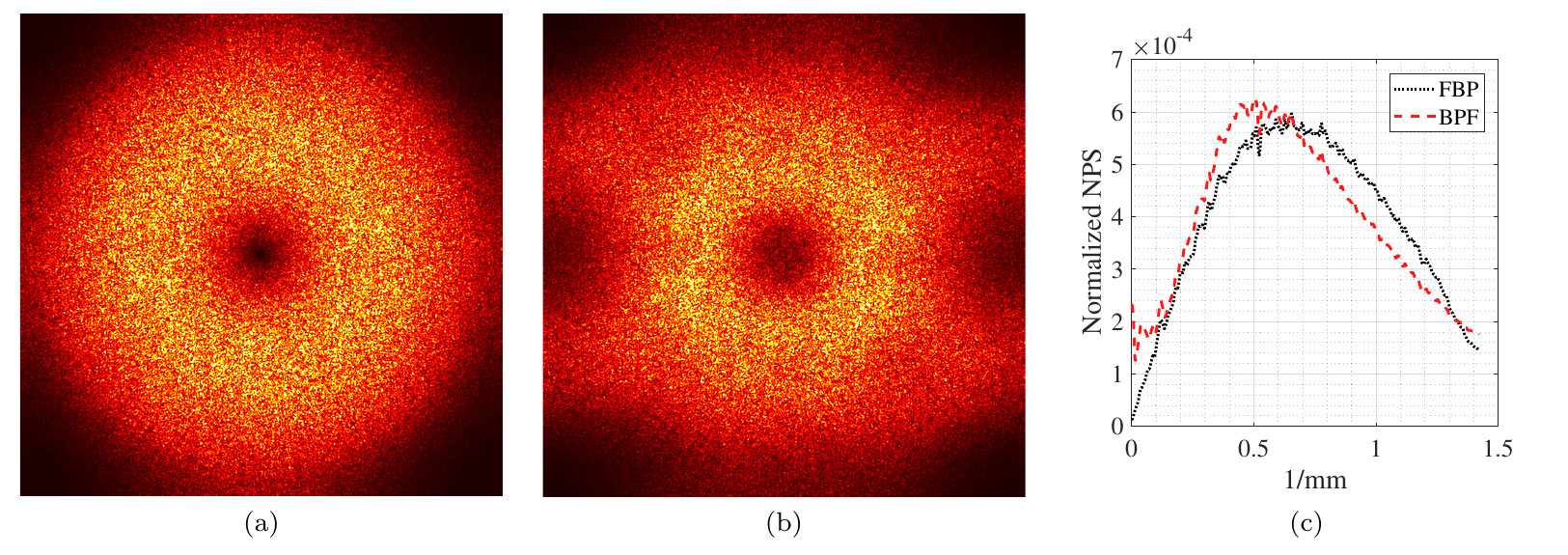}
\vspace{-0.2in}
\caption{The measured NPS results from axial CT images: (a) the 2D NPS map from conventional FBP method, (b) the 2D NPS map from the proposed BPF method, (c) the radially averaged line profile (normalized by the area under the curve).}
\label{fig:nps} 
\end{figure*}

Finally, the CT image noise power spectra (NPS) are analyzed side by side for the two image reconstruction algorithms. Results are shown in Fig.~\ref{fig:nps}. To do so, the central image region with dimension of 256 pixels $\times$ 256 pixels on noise-only image is used to calculate the 2D NPS maps. In total, 50 noise-only samples are used. As can be seen, overall, the proposed BPF CT image reconstruction method is able to generate similar NPS map as the FBP reconstruction method. However, the obtained NPS map from the BPF method contains some interesting features, for example, the four bright legs on the left and right sides of Fig.~\ref{fig:nps}(b). The radially averaged NPS curves also demonstrate the high similarities between the two image reconstruction strategies. Specifically, the NPS profile obtained from the BPF method slightly shifts to the lower frequency part with respect to the FBP method. In addition, the NPS curve from BPF method does not drop to zero at very low frequency region.

\section{Discussion}
In this study, we have demonstrated the feasibility of image-domain laminogram deconvolution BPF-type CT image reconstruction framework using the deep-learning-based technique. Both numerical and experimental results show that the introduced image-domain CNN can reconstruct CT images accurately from the highly blurred laminograms. This proposed CNN network not only being able to generate CT images with similar spatial resolution as from the FBP algorithm, but also helps to maintain the similar noise textures.

\subsection{About the network}
Compared with the domain transform manifold learning CT image reconstruction CNN\cite{zhu2018image}, one major advantage of the CNN-based BPF-type image reconstruction framework is that it can significantly save the computation resource. As pointed in literature\cite{ye2018deep}, the fully-connected layer between the full-size sinogram and full-size CT image may consume a huge amount of GPU resource. This might strongly limit its wide applications in CT image reconstructions, especially for clinical CT images with large pixel dimension, i.e., $512\times512$. As a contrary, our CT image reconstruction network does not extensively require huge GPU resources. This is because the new framework puts this computationally extensive domain-transformation (from sinogram domain to the CT image domain) burden to the laminogram reconstruction procedure, which can be performed analytically and quickly. Once this procedure is finished, the trained CNN can be cascaded afterwards to further reconstruct sharp CT image.

This proposed CNN architecture is inspired by the well-known U-Net architecture\cite{ronneberger2015u}, image debluring network\cite{xu2014deep}, and image super-resolution network\cite{dong2016image, dong2016accelerating}. Our designed network keeps the downsampling and upsampling structures in U-net. However, its middle connection with fully-convolutional layer is replaced by a fully-connected layer. Our experience finds that the fully-connected layer has a better performance in maintaining image sharpness, signal accuracy, as well as the network convergence speed. We also notice that the added shortcut layers are important in maintaining high image spatial resolution, which agrees with the conclusion obtained in previous literature\cite{chen2017low}. We admit that this particular CNN network architecture is purely empirical, and more fundamental work are needed to provide explanations on why and how it works. For instance, it would be an interesting topic to compare the shape of the theoretical deconvolution kernel with the learned approximation by the CNN, from which the effectiveness of the network might be demonstrated by some degree.

\subsection{About the network applications}
During practical applications, the voxel dimension of the laminogram does not have to be the same as used for the training dataset. For example, it is $0.35mm\times0.35mm\times0.35mm$ in this study. For real applications, the laminogram voxel can be reconstructed into any dimension. However, in order to make the trained CNN outputs sharp CT images with correct attenuation coefficient values, the ratio of the trained image voxel dimension to the real voxel dimension needs to be considered. In addition, the real CT imaging geometry also does not have to be the same as for the numerical simulations when generating the training dataset. Actually, the trained CNN result fits well with any type of CT image geometry. This is because the CNN part, which takes care of the deconvolution procedure, is independent of the imaging geometry. Whereas, the procedure of generating laminograms indeed relies on the CT image geometry knowledge.

The current network training dataset is generated from the clinical CT images by converting their Hounsfield Unit into the corresponding linear attenuation coefficient. In order to do this, a empirical conversion factor 0.02/mm, which matches the water attenuation with X-ray beam of about 60 keV, is selected. This is a rough approximation. For CT images scanned with other mean tube potentials, this conversion factor may need to be adjusted correspondingly. To compensate this, we augmented the image values by multiplying an additional factor between 0.50 to 2.00 randomly to each CT image. For current validations, we did not observe any failure. However, it is still important to broaded the data range of the training dataset as wide as possible by adding various clinical CT images to avoid any potential failure in future.

For this discussed CNN, it only accepts input images having a fixed matrix size of $512\times512$. Meanwhile, the output of the CNN also has the same size as of the input image. Clearly, this already trained CNN cannot be used to reconstruct CT images with different sizes, for example, $1024\times1024$ or $256\times256$. In order to make it applicable for varied CT image dimensions, one may need to train this CNN separately regarding to a certain imaging task. Since the training procedure does not take too long, therefore, this should not become a bottleneck for its practical applications.

\subsection{About the future work}
Results show that this novel deconvolution-based BPF-type CT image reconstruction network has the potential to generate similar NPS distributions compared with the FBP reconstruction method. However, for the slight difference between the 2D NPS distribution maps, currently we do not have a clear explanation. Therefore, we would like to study these interesting phenomenons in future.

Moreover, this developed CNN shows potential to reduce image noise compared with the conventional FBP reconstruction results, despite that the training dataset is generated using noise-free images. If training this CNN with dataset containing noise in future, it is very promising that this innovative CNN-based BPF-type CT image reconstruction framework could further reduce CT image noise (and remove CT image artifacts), as have been demonstrated by other recent studies. Therefore, to maintain the image quality while reduce patient dose as low as reasonably achievable (ALARA), we would focus on optimizing the network's dose reduction potential and image artifact removal potential, and eventually generalize the current BPF CT image reconstruction network into a dose-reduction and artifact-free CT image reconstruction network in future. Additionally, results will be compared with other methods as well, including both iterative CT image reconstruction algorithms\cite{stayman2000regularization,elbakri2002statistical, sidky2006accurate,chen2008prior, yu2009compressed, hara2009iterative,yu2011fast} and the pure CNN reconstruction algorithm\cite{zhu2018image}. Finally, the cone-beam CT (CBCT) image reconstruction capability of this CNN-based BPF-type CT reconstruction algorithm will also be investigated as part of the future work.

\section{Conclusions}
In conclusion, we have demonstrated the feasibility of a new BPF CT image reconstruction framework using CNN technique in this work. Within this framework, CT image is reconstructed from the blurred laminogram via a particular CNN learning procedure directly in image-domain. Results show that the CNN not only being able to maintain image spatial resolution and image noise texture, but also has the potential to reduce image noise. Therefore, this proposed CNN-based image-domain BPF-type CT image reconstruction scheme may become a promising alternative for future clinical CT image reconstruction applications.

\section*{Acknowledgment}
We would like to acknowledge Dr. Yaoqin Xie for borrowing us the CT phantoms used in acquiring the experimental validation datasets. We also acknowledge The Cancer Imaging Archive (TCIA) organization for opening access to real clinical CT images. This project is partly supported by Shenzhen Strategic Emerging Industry Development Fund (JSGG20170412100952456).

\bibliography{Bibliography_Paper}

\begin{thebibliography}{33}%
\makeatletter
\providecommand \@ifxundefined [1]{%
 \@ifx{#1\undefined}
}%
\providecommand \@ifnum [1]{%
 \ifnum #1\expandafter \@firstoftwo
 \else \expandafter \@secondoftwo
 \fi
}%
\providecommand \@ifx [1]{%
 \ifx #1\expandafter \@firstoftwo
 \else \expandafter \@secondoftwo
 \fi
}%
\providecommand \natexlab [1]{#1}%
\providecommand \enquote  [1]{``#1''}%
\providecommand \bibnamefont  [1]{#1}%
\providecommand \bibfnamefont [1]{#1}%
\providecommand \citenamefont [1]{#1}%
\providecommand \href@noop [0]{\@secondoftwo}%
\providecommand \href [0]{\begingroup \@sanitize@url \@href}%
\providecommand \@href[1]{\@@startlink{#1}\@@href}%
\providecommand \@@href[1]{\endgroup#1\@@endlink}%
\providecommand \@sanitize@url [0]{\catcode `\\12\catcode `\$12\catcode
  `\&12\catcode `\#12\catcode `\^12\catcode `\_12\catcode `\%12\relax}%
\providecommand \@@startlink[1]{}%
\providecommand \@@endlink[0]{}%
\providecommand \url  [0]{\begingroup\@sanitize@url \@url }%
\providecommand \@url [1]{\endgroup\@href {#1}{\urlprefix }}%
\providecommand \urlprefix  [0]{URL }%
\providecommand \Eprint [0]{\href }%
\providecommand \doibase [0]{http://dx.doi.org/}%
\providecommand \selectlanguage [0]{\@gobble}%
\providecommand \bibinfo  [0]{\@secondoftwo}%
\providecommand \bibfield  [0]{\@secondoftwo}%
\providecommand \translation [1]{[#1]}%
\providecommand \BibitemOpen [0]{}%
\providecommand \bibitemStop [0]{}%
\providecommand \bibitemNoStop [0]{.\EOS\space}%
\providecommand \EOS [0]{\spacefactor3000\relax}%
\providecommand \BibitemShut  [1]{\csname bibitem#1\endcsname}%
\let\auto@bib@innerbib\@empty
\bibitem [{\citenamefont {Kak}\ and\ \citenamefont
  {Slaney}(1988)}]{avinash1988principles}%
  \BibitemOpen
  \bibfield  {author} {\bibinfo {author} {\bibfnamefont {A.~C.}\ \bibnamefont
  {Kak}}\ and\ \bibinfo {author} {\bibfnamefont {M.}~\bibnamefont {Slaney}},\
  }\href@noop {} {\emph {\bibinfo {title} {Principles of computerized
  tomographic imaging}}}\ (\bibinfo  {publisher} {IEEE press},\ \bibinfo {year}
  {1988})\BibitemShut {NoStop}%
\bibitem [{\citenamefont {Hsieh}\ \emph {et~al.}(2009)\citenamefont {Hsieh}
  \emph {et~al.}}]{hsieh2009computed}%
  \BibitemOpen
  \bibfield  {author} {\bibinfo {author} {\bibfnamefont {J.}~\bibnamefont
  {Hsieh}} \emph {et~al.},\ }\bibfield  {title} {\enquote {\bibinfo {title}
  {Computed tomography: principles, design, artifacts, and recent advances},}\
  \ }(\bibinfo {organization} {SPIE Bellingham, WA},\ \bibinfo {year}
  {2009})\BibitemShut {NoStop}%
\bibitem [{\citenamefont {Feldkamp}, \citenamefont {Davis},\ and\ \citenamefont
  {Kress}(1984)}]{feldkamp1984practical}%
  \BibitemOpen
  \bibfield  {author} {\bibinfo {author} {\bibfnamefont {L.}~\bibnamefont
  {Feldkamp}}, \bibinfo {author} {\bibfnamefont {L.}~\bibnamefont {Davis}}, \
  and\ \bibinfo {author} {\bibfnamefont {J.}~\bibnamefont {Kress}},\ }\bibfield
   {title} {\enquote {\bibinfo {title} {Practical cone-beam algorithm},}\
  }\href@noop {} {\bibfield  {journal} {\bibinfo  {journal} {JOSA A}\ }\textbf
  {\bibinfo {volume} {1}},\ \bibinfo {pages} {612--619} (\bibinfo {year}
  {1984})}\BibitemShut {NoStop}%
\bibitem [{\citenamefont {Smith}, \citenamefont {Peters},\ and\ \citenamefont
  {Bates}(1973)}]{smith1973image}%
  \BibitemOpen
  \bibfield  {author} {\bibinfo {author} {\bibfnamefont {P.}~\bibnamefont
  {Smith}}, \bibinfo {author} {\bibfnamefont {T.}~\bibnamefont {Peters}}, \
  and\ \bibinfo {author} {\bibfnamefont {R.}~\bibnamefont {Bates}},\ }\bibfield
   {title} {\enquote {\bibinfo {title} {Image reconstruction from finite
  numbers of projections},}\ }\href@noop {} {\bibfield  {journal} {\bibinfo
  {journal} {Journal of Physics A: Mathematical, Nuclear and General}\ }\textbf
  {\bibinfo {volume} {6}},\ \bibinfo {pages} {361} (\bibinfo {year}
  {1973})}\BibitemShut {NoStop}%
\bibitem [{\citenamefont {Martin}\ \emph {et~al.}(1994)\citenamefont {Martin},
  \citenamefont {Dogan}, \citenamefont {Gormley}, \citenamefont {Knoll},
  \citenamefont {O'Donnell},\ and\ \citenamefont {Wehe}}]{martin1994imaging}%
  \BibitemOpen
  \bibfield  {author} {\bibinfo {author} {\bibfnamefont {J.}~\bibnamefont
  {Martin}}, \bibinfo {author} {\bibfnamefont {N.}~\bibnamefont {Dogan}},
  \bibinfo {author} {\bibfnamefont {J.}~\bibnamefont {Gormley}}, \bibinfo
  {author} {\bibfnamefont {G.}~\bibnamefont {Knoll}}, \bibinfo {author}
  {\bibfnamefont {M.}~\bibnamefont {O'Donnell}}, \ and\ \bibinfo {author}
  {\bibfnamefont {D.}~\bibnamefont {Wehe}},\ }\bibfield  {title} {\enquote
  {\bibinfo {title} {Imaging multi-energy gamma-ray fields with a compton
  scatter camera},}\ }\href@noop {} {\bibfield  {journal} {\bibinfo  {journal}
  {IEEE Transactions on Nuclear Science}\ }\textbf {\bibinfo {volume} {41}},\
  \bibinfo {pages} {1019--1025} (\bibinfo {year} {1994})}\BibitemShut {NoStop}%
\bibitem [{\citenamefont {Danovich}\ and\ \citenamefont
  {Segal}(1994)}]{danovich1994laminogram}%
  \BibitemOpen
  \bibfield  {author} {\bibinfo {author} {\bibfnamefont {Z.}~\bibnamefont
  {Danovich}}\ and\ \bibinfo {author} {\bibfnamefont {Y.}~\bibnamefont
  {Segal}},\ }\bibfield  {title} {\enquote {\bibinfo {title} {Laminogram
  reconstruction through regularizing fourier filtration},}\ }\href@noop {}
  {\bibfield  {journal} {\bibinfo  {journal} {NDT \& E International}\ }\textbf
  {\bibinfo {volume} {27}},\ \bibinfo {pages} {123--130} (\bibinfo {year}
  {1994})}\BibitemShut {NoStop}%
\bibitem [{\citenamefont {Almeida}\ and\ \citenamefont
  {Almeida}(2010)}]{almeida2010blind}%
  \BibitemOpen
  \bibfield  {author} {\bibinfo {author} {\bibfnamefont {M.~S.}\ \bibnamefont
  {Almeida}}\ and\ \bibinfo {author} {\bibfnamefont {L.~B.}\ \bibnamefont
  {Almeida}},\ }\bibfield  {title} {\enquote {\bibinfo {title} {Blind and
  semi-blind deblurring of natural images},}\ }\href@noop {} {\bibfield
  {journal} {\bibinfo  {journal} {IEEE Transactions on Image Processing}\
  }\textbf {\bibinfo {volume} {19}},\ \bibinfo {pages} {36--52} (\bibinfo
  {year} {2010})}\BibitemShut {NoStop}%
\bibitem [{\citenamefont {Campisi}\ and\ \citenamefont
  {Egiazarian}(2017)}]{campisi2017blind}%
  \BibitemOpen
  \bibfield  {author} {\bibinfo {author} {\bibfnamefont {P.}~\bibnamefont
  {Campisi}}\ and\ \bibinfo {author} {\bibfnamefont {K.}~\bibnamefont
  {Egiazarian}},\ }\href@noop {} {\emph {\bibinfo {title} {Blind image
  deconvolution: theory and applications}}}\ (\bibinfo  {publisher} {CRC
  press},\ \bibinfo {year} {2017})\BibitemShut {NoStop}%
\bibitem [{\citenamefont {LeCun}, \citenamefont {Bengio},\ and\ \citenamefont
  {Hinton}(2015)}]{lecun2015deep}%
  \BibitemOpen
  \bibfield  {author} {\bibinfo {author} {\bibfnamefont {Y.}~\bibnamefont
  {LeCun}}, \bibinfo {author} {\bibfnamefont {Y.}~\bibnamefont {Bengio}}, \
  and\ \bibinfo {author} {\bibfnamefont {G.}~\bibnamefont {Hinton}},\
  }\bibfield  {title} {\enquote {\bibinfo {title} {Deep learning},}\
  }\href@noop {} {\bibfield  {journal} {\bibinfo  {journal} {nature}\ }\textbf
  {\bibinfo {volume} {521}},\ \bibinfo {pages} {436} (\bibinfo {year}
  {2015})}\BibitemShut {NoStop}%
\bibitem [{\citenamefont {Jin}\ \emph {et~al.}(2017)\citenamefont {Jin},
  \citenamefont {McCann}, \citenamefont {Froustey},\ and\ \citenamefont
  {Unser}}]{jin2017deep}%
  \BibitemOpen
  \bibfield  {author} {\bibinfo {author} {\bibfnamefont {K.~H.}\ \bibnamefont
  {Jin}}, \bibinfo {author} {\bibfnamefont {M.~T.}\ \bibnamefont {McCann}},
  \bibinfo {author} {\bibfnamefont {E.}~\bibnamefont {Froustey}}, \ and\
  \bibinfo {author} {\bibfnamefont {M.}~\bibnamefont {Unser}},\ }\bibfield
  {title} {\enquote {\bibinfo {title} {Deep convolutional neural network for
  inverse problems in imaging},}\ }\href@noop {} {\bibfield  {journal}
  {\bibinfo  {journal} {IEEE Transactions on Image Processing}\ }\textbf
  {\bibinfo {volume} {26}},\ \bibinfo {pages} {4509--4522} (\bibinfo {year}
  {2017})}\BibitemShut {NoStop}%
\bibitem [{\citenamefont {Chen}\ \emph {et~al.}(2017)\citenamefont {Chen},
  \citenamefont {Zhang}, \citenamefont {Kalra}, \citenamefont {Lin},
  \citenamefont {Chen}, \citenamefont {Liao}, \citenamefont {Zhou},\ and\
  \citenamefont {Wang}}]{chen2017low}%
  \BibitemOpen
  \bibfield  {author} {\bibinfo {author} {\bibfnamefont {H.}~\bibnamefont
  {Chen}}, \bibinfo {author} {\bibfnamefont {Y.}~\bibnamefont {Zhang}},
  \bibinfo {author} {\bibfnamefont {M.~K.}\ \bibnamefont {Kalra}}, \bibinfo
  {author} {\bibfnamefont {F.}~\bibnamefont {Lin}}, \bibinfo {author}
  {\bibfnamefont {Y.}~\bibnamefont {Chen}}, \bibinfo {author} {\bibfnamefont
  {P.}~\bibnamefont {Liao}}, \bibinfo {author} {\bibfnamefont {J.}~\bibnamefont
  {Zhou}}, \ and\ \bibinfo {author} {\bibfnamefont {G.}~\bibnamefont {Wang}},\
  }\bibfield  {title} {\enquote {\bibinfo {title} {Low-dose {CT} with a
  residual encoder-decoder convolutional neural network},}\ }\href@noop {}
  {\bibfield  {journal} {\bibinfo  {journal} {IEEE transactions on medical
  imaging}\ }\textbf {\bibinfo {volume} {36}},\ \bibinfo {pages} {2524--2535}
  (\bibinfo {year} {2017})}\BibitemShut {NoStop}%
\bibitem [{\citenamefont {Wang}\ \emph {et~al.}(2018)\citenamefont {Wang},
  \citenamefont {Ye}, \citenamefont {Mueller},\ and\ \citenamefont
  {Fessler}}]{wang2018image}%
  \BibitemOpen
  \bibfield  {author} {\bibinfo {author} {\bibfnamefont {G.}~\bibnamefont
  {Wang}}, \bibinfo {author} {\bibfnamefont {J.~C.}\ \bibnamefont {Ye}},
  \bibinfo {author} {\bibfnamefont {K.}~\bibnamefont {Mueller}}, \ and\
  \bibinfo {author} {\bibfnamefont {J.~A.}\ \bibnamefont {Fessler}},\
  }\bibfield  {title} {\enquote {\bibinfo {title} {Image reconstruction is a
  new frontier of machine learning.}}\ }\href@noop {} {\bibfield  {journal}
  {\bibinfo  {journal} {IEEE transactions on medical imaging}\ }\textbf
  {\bibinfo {volume} {37}},\ \bibinfo {pages} {1289--1296} (\bibinfo {year}
  {2018})}\BibitemShut {NoStop}%
\bibitem [{\citenamefont {Han}\ and\ \citenamefont
  {Ye}(2018)}]{han2018framing}%
  \BibitemOpen
  \bibfield  {author} {\bibinfo {author} {\bibfnamefont {Y.}~\bibnamefont
  {Han}}\ and\ \bibinfo {author} {\bibfnamefont {J.~C.}\ \bibnamefont {Ye}},\
  }\bibfield  {title} {\enquote {\bibinfo {title} {Framing {U-Net} via deep
  convolutional framelets: Application to sparse-view {CT}},}\ }\href@noop {}
  {\bibfield  {journal} {\bibinfo  {journal} {IEEE transactions on medical
  imaging}\ }\textbf {\bibinfo {volume} {37}},\ \bibinfo {pages} {1418--1429}
  (\bibinfo {year} {2018})}\BibitemShut {NoStop}%
\bibitem [{\citenamefont {Kang}\ \emph {et~al.}(2018)\citenamefont {Kang},
  \citenamefont {Chang}, \citenamefont {Yoo},\ and\ \citenamefont
  {Ye}}]{kang2018deep}%
  \BibitemOpen
  \bibfield  {author} {\bibinfo {author} {\bibfnamefont {E.}~\bibnamefont
  {Kang}}, \bibinfo {author} {\bibfnamefont {W.}~\bibnamefont {Chang}},
  \bibinfo {author} {\bibfnamefont {J.}~\bibnamefont {Yoo}}, \ and\ \bibinfo
  {author} {\bibfnamefont {J.~C.}\ \bibnamefont {Ye}},\ }\bibfield  {title}
  {\enquote {\bibinfo {title} {Deep convolutional framelet denosing for
  low-dose {CT} via wavelet residual network},}\ }\href@noop {} {\bibfield
  {journal} {\bibinfo  {journal} {IEEE transactions on medical imaging}\
  }\textbf {\bibinfo {volume} {37}},\ \bibinfo {pages} {1358--1369} (\bibinfo
  {year} {2018})}\BibitemShut {NoStop}%
\bibitem [{\citenamefont {Yang}\ \emph {et~al.}(2018)\citenamefont {Yang},
  \citenamefont {Yan}, \citenamefont {Zhang}, \citenamefont {Yu}, \citenamefont
  {Shi}, \citenamefont {Mou}, \citenamefont {Kalra}, \citenamefont {Zhang},
  \citenamefont {Sun},\ and\ \citenamefont {Wang}}]{yang2018low}%
  \BibitemOpen
  \bibfield  {author} {\bibinfo {author} {\bibfnamefont {Q.}~\bibnamefont
  {Yang}}, \bibinfo {author} {\bibfnamefont {P.}~\bibnamefont {Yan}}, \bibinfo
  {author} {\bibfnamefont {Y.}~\bibnamefont {Zhang}}, \bibinfo {author}
  {\bibfnamefont {H.}~\bibnamefont {Yu}}, \bibinfo {author} {\bibfnamefont
  {Y.}~\bibnamefont {Shi}}, \bibinfo {author} {\bibfnamefont {X.}~\bibnamefont
  {Mou}}, \bibinfo {author} {\bibfnamefont {M.~K.}\ \bibnamefont {Kalra}},
  \bibinfo {author} {\bibfnamefont {Y.}~\bibnamefont {Zhang}}, \bibinfo
  {author} {\bibfnamefont {L.}~\bibnamefont {Sun}}, \ and\ \bibinfo {author}
  {\bibfnamefont {G.}~\bibnamefont {Wang}},\ }\bibfield  {title} {\enquote
  {\bibinfo {title} {Low dose {CT} image denoising using a generative
  adversarial network with wasserstein distance and perceptual loss},}\
  }\href@noop {} {\bibfield  {journal} {\bibinfo  {journal} {IEEE transactions
  on medical imaging}\ } (\bibinfo {year} {2018})}\BibitemShut {NoStop}%
\bibitem [{\citenamefont {Zhang}\ \emph {et~al.}(2018)\citenamefont {Zhang},
  \citenamefont {Liang}, \citenamefont {Dong}, \citenamefont {Xie},\ and\
  \citenamefont {Cao}}]{zhang2018sparse}%
  \BibitemOpen
  \bibfield  {author} {\bibinfo {author} {\bibfnamefont {Z.}~\bibnamefont
  {Zhang}}, \bibinfo {author} {\bibfnamefont {X.}~\bibnamefont {Liang}},
  \bibinfo {author} {\bibfnamefont {X.}~\bibnamefont {Dong}}, \bibinfo {author}
  {\bibfnamefont {Y.}~\bibnamefont {Xie}}, \ and\ \bibinfo {author}
  {\bibfnamefont {G.}~\bibnamefont {Cao}},\ }\bibfield  {title} {\enquote
  {\bibinfo {title} {A sparse-view ct reconstruction method based on
  combination of densenet and deconvolution.}}\ }\href@noop {} {\bibfield
  {journal} {\bibinfo  {journal} {IEEE transactions on medical imaging}\
  }\textbf {\bibinfo {volume} {37}},\ \bibinfo {pages} {1407--1417} (\bibinfo
  {year} {2018})}\BibitemShut {NoStop}%
\bibitem [{\citenamefont {Zhu}\ \emph {et~al.}(2018)\citenamefont {Zhu},
  \citenamefont {Liu}, \citenamefont {Cauley}, \citenamefont {Rosen},\ and\
  \citenamefont {Rosen}}]{zhu2018image}%
  \BibitemOpen
  \bibfield  {author} {\bibinfo {author} {\bibfnamefont {B.}~\bibnamefont
  {Zhu}}, \bibinfo {author} {\bibfnamefont {J.~Z.}\ \bibnamefont {Liu}},
  \bibinfo {author} {\bibfnamefont {S.~F.}\ \bibnamefont {Cauley}}, \bibinfo
  {author} {\bibfnamefont {B.~R.}\ \bibnamefont {Rosen}}, \ and\ \bibinfo
  {author} {\bibfnamefont {M.~S.}\ \bibnamefont {Rosen}},\ }\bibfield  {title}
  {\enquote {\bibinfo {title} {Image reconstruction by domain-transform
  manifold learning},}\ }\href@noop {} {\bibfield  {journal} {\bibinfo
  {journal} {Nature}\ }\textbf {\bibinfo {volume} {555}},\ \bibinfo {pages}
  {487} (\bibinfo {year} {2018})}\BibitemShut {NoStop}%
\bibitem [{\citenamefont {Gullberg}\ and\ \citenamefont
  {Zeng}(1995)}]{gullberg1995backprojection}%
  \BibitemOpen
  \bibfield  {author} {\bibinfo {author} {\bibfnamefont {G.}~\bibnamefont
  {Gullberg}}\ and\ \bibinfo {author} {\bibfnamefont {G.}~\bibnamefont
  {Zeng}},\ }\bibfield  {title} {\enquote {\bibinfo {title} {Backprojection
  filtering for variable orbit fan-beam tomography},}\ }\href@noop {}
  {\bibfield  {journal} {\bibinfo  {journal} {IEEE transactions on nuclear
  science}\ }\textbf {\bibinfo {volume} {42}},\ \bibinfo {pages} {1257--1266}
  (\bibinfo {year} {1995})}\BibitemShut {NoStop}%
\bibitem [{\citenamefont {He}\ \emph {et~al.}(2016)\citenamefont {He},
  \citenamefont {Zhang}, \citenamefont {Ren},\ and\ \citenamefont
  {Sun}}]{he2016deep}%
  \BibitemOpen
  \bibfield  {author} {\bibinfo {author} {\bibfnamefont {K.}~\bibnamefont
  {He}}, \bibinfo {author} {\bibfnamefont {X.}~\bibnamefont {Zhang}}, \bibinfo
  {author} {\bibfnamefont {S.}~\bibnamefont {Ren}}, \ and\ \bibinfo {author}
  {\bibfnamefont {J.}~\bibnamefont {Sun}},\ }\bibfield  {title} {\enquote
  {\bibinfo {title} {Deep residual learning for image recognition},}\ }in\
  \href@noop {} {\emph {\bibinfo {booktitle} {Proceedings of the IEEE
  conference on computer vision and pattern recognition}}}\ (\bibinfo {year}
  {2016})\ pp.\ \bibinfo {pages} {770--778}\BibitemShut {NoStop}%
\bibitem [{\citenamefont {Kingma}\ and\ \citenamefont
  {Ba}(2014)}]{kingma2014adam}%
  \BibitemOpen
  \bibfield  {author} {\bibinfo {author} {\bibfnamefont {D.~P.}\ \bibnamefont
  {Kingma}}\ and\ \bibinfo {author} {\bibfnamefont {J.}~\bibnamefont {Ba}},\
  }\bibfield  {title} {\enquote {\bibinfo {title} {Adam: A method for
  stochastic optimization},}\ }\href@noop {} {\bibfield  {journal} {\bibinfo
  {journal} {arXiv preprint arXiv:1412.6980}\ } (\bibinfo {year}
  {2014})}\BibitemShut {NoStop}%
\bibitem [{\citenamefont {Ge}\ \emph {et~al.}(2017)\citenamefont {Ge},
  \citenamefont {Ji}, \citenamefont {Zhang}, \citenamefont {Li},\ and\
  \citenamefont {Chen}}]{ge2017k}%
  \BibitemOpen
  \bibfield  {author} {\bibinfo {author} {\bibfnamefont {Y.}~\bibnamefont
  {Ge}}, \bibinfo {author} {\bibfnamefont {X.}~\bibnamefont {Ji}}, \bibinfo
  {author} {\bibfnamefont {R.}~\bibnamefont {Zhang}}, \bibinfo {author}
  {\bibfnamefont {K.}~\bibnamefont {Li}}, \ and\ \bibinfo {author}
  {\bibfnamefont {G.-H.}\ \bibnamefont {Chen}},\ }\bibfield  {title} {\enquote
  {\bibinfo {title} {K-edge energy-based calibration method for photon counting
  detectors},}\ }\href@noop {} {\bibfield  {journal} {\bibinfo  {journal}
  {Physics in Medicine \& Biology}\ }\textbf {\bibinfo {volume} {63}},\
  \bibinfo {pages} {015022} (\bibinfo {year} {2017})}\BibitemShut {NoStop}%
\bibitem [{\citenamefont {Ye}, \citenamefont {Han},\ and\ \citenamefont
  {Cha}(2018)}]{ye2018deep}%
  \BibitemOpen
  \bibfield  {author} {\bibinfo {author} {\bibfnamefont {J.~C.}\ \bibnamefont
  {Ye}}, \bibinfo {author} {\bibfnamefont {Y.}~\bibnamefont {Han}}, \ and\
  \bibinfo {author} {\bibfnamefont {E.}~\bibnamefont {Cha}},\ }\bibfield
  {title} {\enquote {\bibinfo {title} {Deep convolutional framelets: A general
  deep learning framework for inverse problems},}\ }\href@noop {} {\bibfield
  {journal} {\bibinfo  {journal} {SIAM Journal on Imaging Sciences}\ }\textbf
  {\bibinfo {volume} {11}},\ \bibinfo {pages} {991--1048} (\bibinfo {year}
  {2018})}\BibitemShut {NoStop}%
\bibitem [{\citenamefont {Ronneberger}, \citenamefont {Fischer},\ and\
  \citenamefont {Brox}(2015)}]{ronneberger2015u}%
  \BibitemOpen
  \bibfield  {author} {\bibinfo {author} {\bibfnamefont {O.}~\bibnamefont
  {Ronneberger}}, \bibinfo {author} {\bibfnamefont {P.}~\bibnamefont
  {Fischer}}, \ and\ \bibinfo {author} {\bibfnamefont {T.}~\bibnamefont
  {Brox}},\ }\bibfield  {title} {\enquote {\bibinfo {title} {U-net:
  Convolutional networks for biomedical image segmentation},}\ }in\ \href@noop
  {} {\emph {\bibinfo {booktitle} {International Conference on Medical image
  computing and computer-assisted intervention}}}\ (\bibinfo {organization}
  {Springer},\ \bibinfo {year} {2015})\ pp.\ \bibinfo {pages}
  {234--241}\BibitemShut {NoStop}%
\bibitem [{\citenamefont {Xu}\ \emph {et~al.}(2014)\citenamefont {Xu},
  \citenamefont {Ren}, \citenamefont {Liu},\ and\ \citenamefont
  {Jia}}]{xu2014deep}%
  \BibitemOpen
  \bibfield  {author} {\bibinfo {author} {\bibfnamefont {L.}~\bibnamefont
  {Xu}}, \bibinfo {author} {\bibfnamefont {J.~S.}\ \bibnamefont {Ren}},
  \bibinfo {author} {\bibfnamefont {C.}~\bibnamefont {Liu}}, \ and\ \bibinfo
  {author} {\bibfnamefont {J.}~\bibnamefont {Jia}},\ }\bibfield  {title}
  {\enquote {\bibinfo {title} {Deep convolutional neural network for image
  deconvolution},}\ }in\ \href@noop {} {\emph {\bibinfo {booktitle} {Advances
  in Neural Information Processing Systems}}}\ (\bibinfo {year} {2014})\ pp.\
  \bibinfo {pages} {1790--1798}\BibitemShut {NoStop}%
\bibitem [{\citenamefont {Dong}\ \emph {et~al.}(2016)\citenamefont {Dong},
  \citenamefont {Loy}, \citenamefont {He},\ and\ \citenamefont
  {Tang}}]{dong2016image}%
  \BibitemOpen
  \bibfield  {author} {\bibinfo {author} {\bibfnamefont {C.}~\bibnamefont
  {Dong}}, \bibinfo {author} {\bibfnamefont {C.~C.}\ \bibnamefont {Loy}},
  \bibinfo {author} {\bibfnamefont {K.}~\bibnamefont {He}}, \ and\ \bibinfo
  {author} {\bibfnamefont {X.}~\bibnamefont {Tang}},\ }\bibfield  {title}
  {\enquote {\bibinfo {title} {Image super-resolution using deep convolutional
  networks},}\ }\href@noop {} {\bibfield  {journal} {\bibinfo  {journal} {IEEE
  transactions on pattern analysis and machine intelligence}\ }\textbf
  {\bibinfo {volume} {38}},\ \bibinfo {pages} {295--307} (\bibinfo {year}
  {2016})}\BibitemShut {NoStop}%
\bibitem [{\citenamefont {Dong}, \citenamefont {Loy},\ and\ \citenamefont
  {Tang}(2016)}]{dong2016accelerating}%
  \BibitemOpen
  \bibfield  {author} {\bibinfo {author} {\bibfnamefont {C.}~\bibnamefont
  {Dong}}, \bibinfo {author} {\bibfnamefont {C.~C.}\ \bibnamefont {Loy}}, \
  and\ \bibinfo {author} {\bibfnamefont {X.}~\bibnamefont {Tang}},\ }\bibfield
  {title} {\enquote {\bibinfo {title} {Accelerating the super-resolution
  convolutional neural network},}\ }in\ \href@noop {} {\emph {\bibinfo
  {booktitle} {European Conference on Computer Vision}}}\ (\bibinfo
  {organization} {Springer},\ \bibinfo {year} {2016})\ pp.\ \bibinfo {pages}
  {391--407}\BibitemShut {NoStop}%
\bibitem [{\citenamefont {Stayman}\ and\ \citenamefont
  {Fessler}(2000)}]{stayman2000regularization}%
  \BibitemOpen
  \bibfield  {author} {\bibinfo {author} {\bibfnamefont {J.~W.}\ \bibnamefont
  {Stayman}}\ and\ \bibinfo {author} {\bibfnamefont {J.~A.}\ \bibnamefont
  {Fessler}},\ }\bibfield  {title} {\enquote {\bibinfo {title} {Regularization
  for uniform spatial resolution properties in penalized-likelihood image
  reconstruction},}\ }\href@noop {} {\bibfield  {journal} {\bibinfo  {journal}
  {IEEE transactions on medical imaging}\ }\textbf {\bibinfo {volume} {19}},\
  \bibinfo {pages} {601--615} (\bibinfo {year} {2000})}\BibitemShut {NoStop}%
\bibitem [{\citenamefont {Elbakri}, \citenamefont {Fessler}\ \emph
  {et~al.}(2002)\citenamefont {Elbakri}, \citenamefont {Fessler} \emph
  {et~al.}}]{elbakri2002statistical}%
  \BibitemOpen
  \bibfield  {author} {\bibinfo {author} {\bibfnamefont {I.}~\bibnamefont
  {Elbakri}}, \bibinfo {author} {\bibfnamefont {J.}~\bibnamefont {Fessler}},
  \emph {et~al.},\ }\bibfield  {title} {\enquote {\bibinfo {title} {Statistical
  image reconstruction for polyenergetic x-ray computed tomography},}\
  }\href@noop {} {\bibfield  {journal} {\bibinfo  {journal} {Medical Imaging,
  IEEE Transactions on}\ }\textbf {\bibinfo {volume} {21}},\ \bibinfo {pages}
  {89--99} (\bibinfo {year} {2002})}\BibitemShut {NoStop}%
\bibitem [{\citenamefont {Sidky}, \citenamefont {Kao},\ and\ \citenamefont
  {Pan}(2006)}]{sidky2006accurate}%
  \BibitemOpen
  \bibfield  {author} {\bibinfo {author} {\bibfnamefont {E.~Y.}\ \bibnamefont
  {Sidky}}, \bibinfo {author} {\bibfnamefont {C.-M.}\ \bibnamefont {Kao}}, \
  and\ \bibinfo {author} {\bibfnamefont {X.}~\bibnamefont {Pan}},\ }\bibfield
  {title} {\enquote {\bibinfo {title} {Accurate image reconstruction from
  few-views and limited-angle data in divergent-beam {CT}},}\ }\href@noop {}
  {\bibfield  {journal} {\bibinfo  {journal} {Journal of X-ray Science and
  Technology}\ }\textbf {\bibinfo {volume} {14}},\ \bibinfo {pages} {119--139}
  (\bibinfo {year} {2006})}\BibitemShut {NoStop}%
\bibitem [{\citenamefont {Chen}, \citenamefont {Tang},\ and\ \citenamefont
  {Leng}(2008)}]{chen2008prior}%
  \BibitemOpen
  \bibfield  {author} {\bibinfo {author} {\bibfnamefont {G.-H.}\ \bibnamefont
  {Chen}}, \bibinfo {author} {\bibfnamefont {J.}~\bibnamefont {Tang}}, \ and\
  \bibinfo {author} {\bibfnamefont {S.}~\bibnamefont {Leng}},\ }\bibfield
  {title} {\enquote {\bibinfo {title} {Prior image constrained compressed
  sensing ({PICCS}): a method to accurately reconstruct dynamic {CT} images
  from highly undersampled projection data sets},}\ }\href@noop {} {\bibfield
  {journal} {\bibinfo  {journal} {Medical physics}\ }\textbf {\bibinfo {volume}
  {35}},\ \bibinfo {pages} {660--663} (\bibinfo {year} {2008})}\BibitemShut
  {NoStop}%
\bibitem [{\citenamefont {Yu}\ and\ \citenamefont
  {Wang}(2009)}]{yu2009compressed}%
  \BibitemOpen
  \bibfield  {author} {\bibinfo {author} {\bibfnamefont {H.}~\bibnamefont
  {Yu}}\ and\ \bibinfo {author} {\bibfnamefont {G.}~\bibnamefont {Wang}},\
  }\bibfield  {title} {\enquote {\bibinfo {title} {Compressed sensing based
  interior tomography},}\ }\href@noop {} {\bibfield  {journal} {\bibinfo
  {journal} {Physics in medicine \& biology}\ }\textbf {\bibinfo {volume}
  {54}},\ \bibinfo {pages} {2791} (\bibinfo {year} {2009})}\BibitemShut
  {NoStop}%
\bibitem [{\citenamefont {Hara}\ \emph {et~al.}(2009)\citenamefont {Hara},
  \citenamefont {Paden}, \citenamefont {Silva}, \citenamefont {Kujak},
  \citenamefont {Lawder},\ and\ \citenamefont {Pavlicek}}]{hara2009iterative}%
  \BibitemOpen
  \bibfield  {author} {\bibinfo {author} {\bibfnamefont {A.~K.}\ \bibnamefont
  {Hara}}, \bibinfo {author} {\bibfnamefont {R.~G.}\ \bibnamefont {Paden}},
  \bibinfo {author} {\bibfnamefont {A.~C.}\ \bibnamefont {Silva}}, \bibinfo
  {author} {\bibfnamefont {J.~L.}\ \bibnamefont {Kujak}}, \bibinfo {author}
  {\bibfnamefont {H.~J.}\ \bibnamefont {Lawder}}, \ and\ \bibinfo {author}
  {\bibfnamefont {W.}~\bibnamefont {Pavlicek}},\ }\bibfield  {title} {\enquote
  {\bibinfo {title} {Iterative reconstruction technique for reducing body
  radiation dose at {CT}: feasibility study},}\ }\href@noop {} {\bibfield
  {journal} {\bibinfo  {journal} {American Journal of Roentgenology}\ }\textbf
  {\bibinfo {volume} {193}},\ \bibinfo {pages} {764--771} (\bibinfo {year}
  {2009})}\BibitemShut {NoStop}%
\bibitem [{\citenamefont {Yu}\ \emph {et~al.}(2011)\citenamefont {Yu},
  \citenamefont {Thibault}, \citenamefont {Bouman}, \citenamefont {Sauer},\
  and\ \citenamefont {Hsieh}}]{yu2011fast}%
  \BibitemOpen
  \bibfield  {author} {\bibinfo {author} {\bibfnamefont {Z.}~\bibnamefont
  {Yu}}, \bibinfo {author} {\bibfnamefont {J.-B.}\ \bibnamefont {Thibault}},
  \bibinfo {author} {\bibfnamefont {C.~A.}\ \bibnamefont {Bouman}}, \bibinfo
  {author} {\bibfnamefont {K.~D.}\ \bibnamefont {Sauer}}, \ and\ \bibinfo
  {author} {\bibfnamefont {J.}~\bibnamefont {Hsieh}},\ }\bibfield  {title}
  {\enquote {\bibinfo {title} {Fast model-based x-ray {CT} reconstruction using
  spatially nonhomogeneous {ICD} optimization},}\ }\href@noop {} {\bibfield
  {journal} {\bibinfo  {journal} {IEEE Transactions on image processing}\
  }\textbf {\bibinfo {volume} {20}},\ \bibinfo {pages} {161--175} (\bibinfo
  {year} {2011})}\BibitemShut {NoStop}%
\end{thebibliography}%

\end{document}